\documentclass[review]{elsarticle}
\usepackage[margin=2cm]{geometry}
\usepackage[misc, geometry]{ifsym} 
\usepackage[T1]{fontenc}
\usepackage[utf8]{inputenc}
\usepackage{textgreek} 
\RequirePackage{amssymb, amsmath, amsfonts}

\newtheorem{definition}{Definition}

\usepackage[x11names, usenames, dvipsnames, svgnames, table]{xcolor}
\usepackage{enumitem}

\usepackage[skip=2pt, font=footnotesize,labelfont=bf]{caption}
\RequirePackage[skip=1.75pt, font=footnotesize, labelfont=sf, textfont=sf]{subcaption}
\usepackage{textcomp}
\usepackage{tabularx}{\tiny }
\usepackage{xtab, booktabs} 
\usepackage{lscape}
\usepackage{longtable}
\usepackage{adjustbox}
\usepackage{setspace}
\usepackage{multirow}
\usepackage{multicol}
\usepackage{tcolorbox}
\usepackage{hhline}
\usepackage{colortbl}
\usepackage{rotating}
\usepackage[normalem]{ulem}
\usepackage{bm}
\usepackage{nicefrac}  
\usepackage{microtype}  
 \usepackage[hyphens]{url}
\usepackage[colorlinks]{hyperref}
\AtBeginDocument{%
\hypersetup{
    citecolor=Purple,
    linkcolor=OliveGreen,
    urlcolor=NavyBlue,
    }
}

\RequirePackage{algorithm}
\RequirePackage{algorithmic}
\usepackage[algo2e, ruled, linesnumbered, resetcount]{algorithm2e}
\def\HiLi{\leavevmode\rlap{\hbox to \hsize{\color{yellow!50}\leaders\hrule height .80\baselineskip depth .8ex\hfill}}}

 \journal{Arxiv}

\begin{document}
    
\begin{frontmatter}
\title{\textbf{Natural Language Requirements Testability Measurement Based on Requirement Smells}}
\author[firstaddress]{Morteza Zakeri-Nasrabadi}
\ead{morteza\_zakeri@comp.iust.ac.ir}

\author[firstaddress]{Saeed Parsa\corref{correspondingauthor}}
\cortext[correspondingauthor]{Corresponding author}
\ead{parsa@iust.ac.ir}

\address[firstaddress]{School of Computer Engineering, Iran University of Science and Technology, Hengam St., Resalat Square, Tehran, Tehran, Iran.}

\begin{abstract}
Requirements form the basis for defining software systems' obligations and tasks. Testable requirements help prevent failures, reduce maintenance costs, and make it easier to perform acceptance tests. However, despite the importance of measuring and quantifying requirements testability, no automatic approach for measuring requirements testability has been proposed based on the requirements smells, which are at odds with the requirements testability. This paper presents a mathematical model to evaluate and rank the natural language requirements testability based on an extensive set of nine requirements smells, detected automatically, and acceptance test efforts determined by requirement length and its application domain. Most of the smells stem from uncountable adjectives, context-sensitive, and ambiguous words. A comprehensive dictionary is required to detect such words. We offer a neural word-embedding technique to generate such a dictionary automatically. Using the dictionary, we could automatically detect Polysemy smell (domain-specific ambiguity) for the first time in 10 application domains. Our empirical study on nearly 1000 software requirements from six well-known industrial and academic projects demonstrates that the proposed smell detection approach outperforms Smella, a state-of-the-art tool, in detecting requirements smells. The precision and recall of smell detection are improved with an average of 0.03 and 0.33, respectively, compared to the state-of-the-art. The proposed requirement testability model measures the testability of 985 requirements with a mean absolute error of 0.12 and a mean squared error of 0.03, demonstrating the model's potential for practical use.
\end{abstract}
\end{frontmatter}

\section{Introduction}\label{sec:introduction}
\noindent
Requirements are the foundation upon which software systems are built. A minor deviation from requirements is strictly a failure that may cause fundamental damages to the ultimate software product. Many failures in software systems stem from poor requirements definition \cite{DosSantos2018}. Correctly understanding what the system must do is critical to its success and validity \cite{Femmer2019}. 
Requirements should be tested in different stages of the software development life cycle (SDLC) to ensure they are understood and met appropriately.

Requirements testability is the degree to which requirements can be tested \cite{Hayes2015}. ISO/IEC/IEEE 24765 standard defines requirements testability as the "extent to which an objective and feasible test can be designed to determine whether a requirement is met" \cite{ISO/IEC/IEEE20171}.
 Unfortunately, few studies have focused on measuring and quantifying requirements testability \cite{Garousi2019}. 
 The root cause of software failure is the misunderstanding of the requirements due to the requirements smell. However, it is observed that instead of requirements testability, researchers have focused on source code testability \cite{Khan2009, Shaheen2014, Terragni2020, Nasrabadi2021, Zakeri20221}, which is itself affected by the testability of the requirements. In other words, instead of addressing the cause, they have focused on resolving the symptoms.
In functional and acceptance testing, requirements are considered obligatory artifacts, conducting the tests towards the users' expectations of the software \cite{Ammann2016, Beer2017}.
Most of the proposed techniques for measuring requirements testability are domain-specific \cite{Izosimov2012}, semi-automated \cite{Genova2013}, or based on a too-small data set for reliable machine learning training \cite{Hayes2015}, making them inappropriate to use in most real-world applications. 

Requirements testability is automatically measurable, provided that any smells can be detected and identified in the requirement definition. A \emph{requirement smell} is a quality issue that can be improved by modifying the requirement definition via removing any ambiguity, misunderstandings, or difficulties with the validation and design of acceptance tests. 
Femmer et al.  \cite{Femmer2014, Femmer20172} have introduced nine requirement smells and created a tool, Smella, to detect eight of the nine smells. Smella utilizes natural language processing (NLP) techniques to detect four out of the eight smells. 
The utilized NLP techniques are part of speech (POS) tagging \cite{Jurafsky2009, Slav2011}, lemmatization \cite{Jurafsky2009}, and morphological analysis \cite{Manning1999}. 
It recognizes the other four smells, including subjective language, ambiguous adverbs and adjectives, loopholes, and non-verifiable terms, by simply searching the requirement's words in a predefined dictionary of the smelly words. Their dictionary has been created manually by software engineering experts.

Most automated techniques for detecting ambiguity in requirements definitions rely on a predefined dictionary of ambiguous words \cite{Wilson1997, Fabbrini2001, Tjong2013, Femmer20172}. These dictionaries include words that are generally regarded as ambiguous \cite{Gleich2010, AmbiguityHandbook2003}. For instance, two requirements analysis tools, Quality Analyzer of Requirements Specifications (QuARS) \cite{Fabbrini2001} and and Systemized Requirements Engineering Environment (SREE) \cite{Tjong2013} use domain-independent dictionaries, including vague words in every domain. Hence, the degree of detected smells by these tools is somewhat context-independent. However, some words are not vague, but their definitions vary depending on how they are used. 
The ones most difficult to handle are the \emph{polysemous} words having different meanings from computer terminology in domains other than computer science. Considering the impacts of context-sensitive and vague words on misunderstanding and, consequently, reducing the testability of requirements, a comprehensive dictionary including both ambiguous and context-sensitive words could be promising.

We found it challenging to prepare a suitable list of smelly words for all application domains, and there is no standard and comprehensive dictionary with such a purpose. Software requirements may be expressed in various languages other than English. Therefore, a systematic and language-agnostic approach is required to create a desirable dictionary of smelly words, along with their smell types.
The only dictionary including context-sensitive nouns in five different domains is introduced in  \cite{Ferrari2017}. Ferrari et al. \cite{Ferrari2017, Ferrari2018, Ferrari2019} have proposed a method based on \emph{word embedding} to automatically detect domain-specific nouns by comparing their cosine similarities in five different domains. They have only focused on nouns, while other parts of speech such as verbs, adverbs, and adjectives may cause ambiguity. We extend their dictionary to include smelly words rather than solely nouns in ten different domains. Our automatically generated dictionary is of great use to determine the clarity and testability of textual requirement artifacts.

In this paper, we define requirements testability as a function of requirement smells and quantify the value of testability for any given requirement without imposing any template \cite{Arora2015} for requirement definitions or restrictions on the application domain \cite{Izosimov2012}. To this aim, a tool for automatically detecting nine requirement smells, boosted by a comprehensive dictionary of smelly words, is proposed. 
To create such a dictionary, inspired by \cite{Ferrari2017}, we identified the 1,000 most frequently used words, common in the computer science domain and ten other application domains. The different meanings of each frequent word were extracted using word embedding and vector similarity on a large corpus of Wikipedia documents. In summary, the following contributions are made by this study:

\begin{enumerate}[label=\roman*., noitemsep, labelwidth=!, labelindent=0pt, topsep=0pt]
\item{
To offer a new mathematical model, defining requirements testability in terms of the requirements smells, size of the requirements, and application domain. Our empirical studies, described in Section \ref{sec:requirement-testability-analysis}, demonstrate the correctness and applicability of the proposed requirements testability measurement equation.
}

\item{
To automatically generate a dictionary of \emph{smelly} or \emph{context-sensitive} words having different meanings in different application domains and the word's smell types. A neural Word2Vec model is trained to calculate the cosine similarity between frequent words commonly used in computer science and ten other application domains. 
}

\item{To introduce two new types of requirement smells, called \emph{Polysemy} and \emph{Uncertain Verbs}. The definitions are given in Section \ref{Requirement-smells}.}

\item{
To introduce the first publicly available requirements testability dataset, including requirements samples, their smells, and testability measures. The dataset has been manually annotated and validated by requirements engineer experts for natural language requirements in English. The annotated requirements dataset is available at \href{https://doi.org/10.5281/zenodo.4266727}{\emph{https://doi.org/10.5281/zenodo.4266727}}.
}
\end{enumerate}

The proposed approach is supported by a web-based requirements smell detection tool that automatically highlights smelly words and determines the testability of each input requirement. Our tool named Automatic Requirements Testability Analyzer (ARTA) is a web application with a rich graphical user interface. The application receives a requirement as input, determines the smelly words visually, and reports its testability value. Researchers and practitioners can define a project, import all project requirements from an Excel file, and manually specify smelly words for each requirement. A web-based tool facilitates creating and managing large requirements datasets, while it can be used to service software developers, testers, and stakeholders to meet high-quality requirements. The ARTA tool is publicly available at \href{https://m-zakeri.github.io/ARTA}{\emph{https://m-zakeri.github.io/ARTA}}.

The remaining parts of this paper are organized as follows. In Section \ref{sec:related-work}, the related works are discussed. Section \ref{sec:methodology} explains our proposed method for measuring software requirements testability. Section \ref{sec:evaluation} evaluates and discusses the results of our experiments with the proposed method. Threats to validity are discussed in Section \ref{sec:threats-to-validity}. Finally, Section \ref{sec:conclusion} concludes the paper and outlines future works.

\section{Related work}\label{sec:related-work}
\noindent
There have been significant advances in studying ambiguities in natural language requirements specifications as a basis for evaluating the quality of the requirements \cite{Zhao2021}. However, in total, we could find only a few studies explicitly on requirements testability \cite{Izosimov2012, Hayes2015}. Despite several standards by IEEE, ISO, and IET \cite{IEEE1990, ISO/IEC/IEEE20171, ISO/IEC/IEEE20172, ISO/IEC2018} defining software requirements testability, there has been no major intention to quantify the requirements testability.

Table \ref{table:1} summarizes the related works in terms of their topics, techniques, evaluations, key advantages, and disadvantages. Considering the column labeled "Approach/ Technique," it is observed that most of these approaches \cite{Wilson1997, Fabbrini2001, Gleich2010, Genova2013, Femmer20172} evaluate a given requirement text by looking up the words in a dictionary of smelly words. Moreover, this column shows that dictionaries are mainly built manually by experts and researchers. 
The other observation, inspired by the last column, is that most dictionaries contain inherently ambiguous words \cite{Wilson1997, Gleich2010, Femmer20172}. They do not include domain-specific and context-sensitive ambiguous words. 
Column 2, labeled "Topic," shows that despite the strict emphasis on the potential impact of ambiguity on testability, defined by several standards, there are only a few approaches \cite{Wilson1997, Hayes2015} directly addressing the measurement of requirements testability. The column, labeled "Evaluation," shows that most approaches provide Precision and Recall. However, their datasets are not available.

\begin{table}[!h]
    \centering
    \caption{Summary and comparison of semi/ fully automated requirement quality assurance approaches}
    \label{table:1}
    \resizebox{1.0\linewidth}{!}{%
        \begin{tabular}{llllll} 
            \hline
            \rowcolor[rgb]{1,0.886,0.682} \begin{tabular}[c]{@{}>{\cellcolor[rgb]{1,0.886,0.682}}l@{}} Tool and \\Publications \end{tabular} & Topic                                                                                                                          & Approach/ Technique                                                                                                                                                 & Evaluation                                                                                                                                & Key advantages                                                                                                                                                     & Key disadvantages                                                                                                                                                                                              \\ 
            \hline
            \rowcolor[rgb]{0.929,0.929,0.929} \begin{tabular}[c]{@{}>{\cellcolor[rgb]{0.929,0.929,0.929}}l@{}} ARM \\ \cite{Wilson1997} \end{tabular}        & \begin{tabular}[c]{@{}>{\cellcolor[rgb]{0.929,0.929,0.929}}l@{}} Requirement quality \\attributes and indicators \end{tabular} & \begin{tabular}[c]{@{}>{\cellcolor[rgb]{0.929,0.929,0.929}}l@{}} - Manual dictionary \\- size metrics \\- readability metrics  ~ \end{tabular}                      & \begin{tabular}[c]{@{}>{\cellcolor[rgb]{0.929,0.929,0.929}}l@{}} Visualization and\\ interpretation  \end{tabular}                        & \begin{tabular}[c]{@{}>{\cellcolor[rgb]{0.929,0.929,0.929}}l@{}} Considering different \\quality aspects \end{tabular}                                             & \begin{tabular}[c]{@{}>{\cellcolor[rgb]{0.929,0.929,0.929}}l@{}} - Limited dictionary of \\general ambiguous words \\- Weak empirical \\evaluation \end{tabular}                                               \\
            \begin{tabular}[c]{@{}l@{}} QuARS \\ \cite{Fabbrini2001} \\ \cite{Lami2004} \end{tabular}                                                                       & \begin{tabular}[c]{@{}l@{}} Quality model for \\requirements \end{tabular}                                                     & \begin{tabular}[c]{@{}l@{}} - Manual dictionary \\- syntactical analyzer  ~ \end{tabular}                                                                           & \begin{tabular}[c]{@{}l@{}} - Real examples \\- Quantification \end{tabular}                                                              & \begin{tabular}[c]{@{}l@{}} Modular design for \\multi-lingual usage \end{tabular}                                                                                 & \begin{tabular}[c]{@{}l@{}} - Limited dictionary of \\general ambiguous words \\- Weak empirical \\evaluation \end{tabular}                                                                                    \\
            \rowcolor[rgb]{0.929,0.929,0.929} \cite{Gleich2010}                                                                                           & Ambiguity detection                                                                                                            & \begin{tabular}[c]{@{}>{\cellcolor[rgb]{0.929,0.929,0.929}}l@{}} - Manual dictionary \\(Ambiguity Handbook) \\- POS tagging \\-Regular expressions  ~ \end{tabular} & \begin{tabular}[c]{@{}>{\cellcolor[rgb]{0.929,0.929,0.929}}l@{}} - Precision: \\34-97\% \\ - Recall: \\53-86\% \end{tabular}              & \begin{tabular}[c]{@{}>{\cellcolor[rgb]{0.929,0.929,0.929}}l@{}} - More comprehensive \\dictionary \\- Strong empirical \\evaluation \end{tabular}                 & \begin{tabular}[c]{@{}>{\cellcolor[rgb]{0.929,0.929,0.929}}l@{}} - Only detect general \\ambiguities \\ - The tool is not available \end{tabular}                                                                     \\
            \cite{Yang2011}                                                                                                                              & \begin{tabular}[c]{@{}l@{}} Ambiguity detection \\(Anaphoric analysis) \end{tabular}                                           & \begin{tabular}[c]{@{}l@{}} Machine learning \\ (classification)  ~ \end{tabular}                                                                                   & \begin{tabular}[c]{@{}l@{}} - Accuracy: \\76.25\% \end{tabular}                                                                           &  \begin{tabular}[c]{@{}l@{}}  Strong feature \\ engineering \end{tabular}                                                                                                               & Small training set                                                                                                                                                                                             \\
            \rowcolor[rgb]{0.929,0.929,0.929} \begin{tabular}[c]{@{}>{\cellcolor[rgb]{0.929,0.929,0.929}}l@{}} SREE \\ \cite{Tjong2013} \end{tabular}       & Ambiguity detection                                                                                                            & \begin{tabular}[c]{@{}>{\cellcolor[rgb]{0.929,0.929,0.929}}l@{}} Manual dictionary \\(Ambiguity Indicator \\Corpus) \end{tabular}                                   & \begin{tabular}[c]{@{}>{\cellcolor[rgb]{0.929,0.929,0.929}}l@{}} - Precision: \\68\% \\ - Recall: \\ $\approx$ 100\% \end{tabular} & 
            Very high (robust) recall
            & Relativity low precision                                                                                                                                                                                       \\
            \cite{Ferrari20182}                                                                                                                               & Ambiguity detection                                                                                                            & \begin{tabular}[c]{@{}l@{}} - SREE \\ \cite{Tjong2013} ~ \\- Regular expressions \end{tabular}                                                                                       & \begin{tabular}[c]{@{}l@{}} - Precision: \\83.16\% \\- Recall: \\85.39\%  ~ \end{tabular}                                                 & \begin{tabular}[c]{@{}l@{}} Strong empirical \\evaluation \end{tabular}                                                                                            & \begin{tabular}[c]{@{}l@{}} Limited to requirements \\ in the railway domain \end{tabular}                                                                                                                          \\
            \rowcolor[rgb]{0.929,0.929,0.929} \begin{tabular}[c]{@{}>{\cellcolor[rgb]{0.929,0.929,0.929}}l@{}} RQA \\ \cite{Genova2013} \end{tabular}         & \begin{tabular}[c]{@{}>{\cellcolor[rgb]{0.929,0.929,0.929}}l@{}} Requirement quality \\attributes and indicators \end{tabular} & \begin{tabular}[c]{@{}>{\cellcolor[rgb]{0.929,0.929,0.929}}l@{}} - Manual dictionary \\(Forbidden Terms) \\ - Step function  ~ \end{tabular}                        & -                                                                                                                                         & \begin{tabular}[c]{@{}>{\cellcolor[rgb]{0.929,0.929,0.929}}l@{}} Commercial tool \\supported \end{tabular}                                                         & No empirical evaluation                                                                                                                                                                                        \\
            \cite{Hayes2015}                                                                                                      & Testability prediction                                                                                                         & \begin{tabular}[c]{@{}l@{}} Machine learning \\(classification)  ~ \end{tabular}                                                                                    & \begin{tabular}[c]{@{}l@{}} - Precision: \\98\% \\- Recall: \\97\% \\ - ROC: \\99\% \end{tabular}                                         & \begin{tabular}[c]{@{}l@{}} A fully automated \\approach for \\testability measurement  \end{tabular}                                                              & \begin{tabular}[c]{@{}l@{}} A very small dataset with \\89 requirement samples  \end{tabular}                                                                                                                  \\
            \rowcolor[rgb]{0.929,0.929,0.929} \begin{tabular}[c]{@{}>{\cellcolor[rgb]{0.929,0.929,0.929}}l@{}} Smella \\ \cite{Femmer20172} \end{tabular}  & \begin{tabular}[c]{@{}>{\cellcolor[rgb]{0.929,0.929,0.929}}l@{}} Ambiguity detection \\(Requirements Smells) \end{tabular}     & \begin{tabular}[c]{@{}>{\cellcolor[rgb]{0.929,0.929,0.929}}l@{}} - Manual dictionary \\- Lemmatization \\ - POS tagging \end{tabular}                               & \begin{tabular}[c]{@{}>{\cellcolor[rgb]{0.929,0.929,0.929}}l@{}} - Precision: \\48\% \\- Recall: \\87\% \end{tabular}                     & \begin{tabular}[c]{@{}>{\cellcolor[rgb]{0.929,0.929,0.929}}l@{}} - Follows ISO/IEC/IEEE \\29148:2011 \\standard \\- Strong empirical \\evaluation \end{tabular} & \begin{tabular}[c]{@{}>{\cellcolor[rgb]{0.929,0.929,0.929}}l@{}} - No public tool, \\and datasets  ~ \end{tabular}                                                                                             \\
            \begin{tabular}[c]{@{}l@{}} \cite{Ferrari2017} \\ \cite{Ferrari2018} \\ \cite{Ferrari2019} \end{tabular}                                                                           & \begin{tabular}[c]{@{}l@{}} Ambiguity detection \\ (Cross-domain) \end{tabular}                                                & Word-embedding                                                                                                                                                      & \begin{tabular}[c]{@{}l@{}} - Kendall’s Tau: \\88\% \\- Real examples  \end{tabular}                                                      & \begin{tabular}[c]{@{}l@{}} Detecting cross-domain \\ambiguity fully \\automated \end{tabular}                                                                     & \begin{tabular}[c]{@{}l@{}} Only detects ambiguous \\nouns in five domains \end{tabular}                                                                                                                       \\
            \rowcolor[rgb]{0.929,0.929,0.929} \begin{tabular}[c]{@{}>{\cellcolor[rgb]{0.929,0.929,0.929}}l@{}} REVV-Light \\ \cite{Dalpiaz2019} \end{tabular} & \begin{tabular}[c]{@{}>{\cellcolor[rgb]{0.929,0.929,0.929}}l@{}} Ambiguity detection \\(Near-synonyms) \end{tabular}           & \begin{tabular}[c]{@{}>{\cellcolor[rgb]{0.929,0.929,0.929}}l@{}} Semantic Folding \\Theory \\ \cite{Webber2015} \end{tabular}                                                       & \begin{tabular}[c]{@{}>{\cellcolor[rgb]{0.929,0.929,0.929}}l@{}} - Precision: \\51\% \\- Recall: \\25\% \end{tabular}                     & \begin{tabular}[c]{@{}>{\cellcolor[rgb]{0.929,0.929,0.929}}l@{}} Open-source tool \\supported for analyzing \\and visualizing \\ user-stories \end{tabular}           & \begin{tabular}[c]{@{}>{\cellcolor[rgb]{0.929,0.929,0.929}}l@{}} - A semi-automated \\method required manual \\assessment \\- Only works on user \\stories written in a specific\\ template \end{tabular}  \\
             \begin{tabular}[c]{@{}>{\cellcolor[rgb]{1,1,1}}l@{}}  \cite{Ezzini2021} \end{tabular} & \begin{tabular}[c]{@{}>{\cellcolor[rgb]{1,1,1}}l@{}} Ambiguity detection \\(Coordination ambiguity \\ and prepositional-phrase \\ attachment ambiguity) \end{tabular}           & \begin{tabular}[c]{@{}>{\cellcolor[rgb]{1,1,1}}l@{}}  - POS tagging \\ - Rule set  \\- Keywords extraction  \end{tabular}                                                       & \begin{tabular}[c]{@{}>{\cellcolor[rgb]{1,1,1}}l@{}} - Precision: \\80\% \\- Recall: \\89\% \end{tabular}                     & \begin{tabular}[c]{@{}>{\cellcolor[rgb]{1,1,1}}l@{}} Low false positive rate \end{tabular}           & \begin{tabular}[c]{@{}>{\cellcolor[rgb]{1,1,1}}l@{}} - Manually defined patterns \\- Low coverage of patterns \end{tabular}  \\
            \hline
        \end{tabular}
    }
\end{table}

The ambiguity detection method, called Automated Requirements Measurement (ARM) \cite{Wilson1997}, had been applied to 46 requirement specification files for the software assurance technology center (STAC) projects in National Aeronautics and Space Administration (NASA). However, they have not provided any report on the Precision and Recall of their evaluation results. Their report is only a quantitative overview of their evaluation results on 46 specifications from NASA.

Another semi-automated approach, called Quality Analyzer of Requirements Specifications (QuARS) \cite{Fabbrini2001, Lami2004}, has been applied to requirement documents taken from four industrial projects. They have reported the number of selected defects in the requirements definitions. Their dictionaries only contain inherently ambiguous words in the English language, and they have not evaluated their system in terms of Precision and Recall. The QuARS tool has been recently applied to detect variability in natural language requirement documents \cite{Fantechi2019}.

Chantree et al. \cite{Chantree2006} have proposed a technique that automatically alerts authors of requirements to the presence of nocuous ambiguities based on four different heuristics and their combinations. 
Their heuristics includes coordination matches, distributional similarity, collocation frequency, and morphology. However, they have focused on detecting nocuous ambiguities, confirmed by multiple readers in multi-readings.

Gleich et al. \cite{Gleich2010} have automated the detection of six types of ambiguity using guidelines for ambiguity detection defined in the Ambiguity Handbook \cite{AmbiguityHandbook2003}. Their tool uses manually predefined dictionaries and light NLP techniques such as POS tagging to detect ambiguous words. It has been evaluated on 50 German and 50 English sentences with known ambiguities marked by experts. The reported precision varies between 34\% for the pure experts' opinion to 97\% for a more guideline-based gold standard. The relation between ambiguity and requirement quality attributes, \emph{e.g.}, testability, has not been discussed. 

Yang et al. \cite{Yang2011} have studied anaphoric ambiguity, which occurs when readers may disagree on how pronouns should be interpreted. They have built some machine learning models to classify instances of ambiguity into nocuous or innocuous based on a set of 17 features made up of their heuristics scores. A set of 200 anaphoric instances manually labeled by experts have been used to evaluate their learned models' performance. An accuracy of 76.25\% has been reported for their best model. Their dataset, containing only 200 instances, seems to be very small for detecting such ambiguities. A larger dataset may increase their model accuracy.

Systemized Requirements Engineering Environment (SREE) \cite{Tjong2013} is another ambiguity-finding tool (AFT) based on manually collected dictionaries of ambiguous words. SREE aims at the detection of ambiguities with a Recall of 100\%. To achieve this, the authors have avoided using any NLP techniques, and instead, they have collected a large dictionary of ambiguous words called Ambiguity Indicator Corpus (AIC). A careless increase in Recall has led to a pretty low Precision of 68\%, as reported by SREE's creators, while they could not achieve a 100\% recall at all. In addition, SREE's AIC does not include any polysemy words.

Ferarri et al. \cite{Ferrari20182} have combined SREE \cite{Tjong2013} with a set of manually defined language structure patterns to find requirement defects in the railway domain. After several refinements of their patterns, they could achieve a Precision of 83.16\% and a Recall of 85.39\% on a dataset of 1866 railway requirements. Therefore, the obtained patterns are likely to depend on their dataset.

G{\'{e}}nova et al. \cite{Genova2013} have used a list of 11 desirable properties, three of which are understandability, unambiguity, and atomicity. They used their own metrics and threshold to quantify and evaluate these properties for textual requirements. However, manual determination of the thresholds to rank requirements definitions without any empirical evaluations is not generally reliable. 

Hayes et al. \cite{Hayes2015} have used logistic regression to predict requirements testability based on several measures, including the number of predicates, number of words in each requirement, and number of complex words, by manually labeling requirements as testable and non-testable. They have used a dataset consisting of 89 requirements extracted from two projects to implement their approach. However, 89 samples cannot provide an accurate machine-learning prediction model.

Femmer et al. \cite{Femmer2014, Femmer20172} have introduced the notion of requirement smell in the context of requirements engineering. They have offered a requirement evaluation tool, Smella, which can detect eight different smells in requirements written in the German language. Their hand-made dictionary could detect four types of these smells. We have extended their lists of smells with two newly defined smells and have proposed a new approach to the automatic construction of a dictionary of smelly words. 

Ferrari et al.  \cite{Ferrari2017, Ferrari2018, Ferrari2019} have proposed a method based on word embedding \cite{Young2018} to detect domain-specific ambiguities in software requirements. They determine the similarity of typical computer science nouns, \emph{e.g.}, "interface" and "database," when used in five different domains. 
They have only focused on nouns, while other parts of speech, such as verbs, adverbs, and adjectives, may cause ambiguity in the requirements definition. We extend their approach to include ten different application domains. We also include different parts of speech available in user requirements to create a comprehensive dictionary for detecting various smells and then measure requirements testability. 

Dalpiaz et al. \cite{Dalpiaz2019} have blended information visualization with NLP techniques, conceptual model extraction, and semantic similarity to find near-synonym terms in user stories. They specifically search for a type of terminological ambiguity, called Near-synonyms or Plesionyms, which are distinct terms referring to the same denotation. Their approach is limited to user stories with predefined templates. Their results reject the hypotheses that automatic ambiguity detection outperforms manual inspection in terms of Precision and Recall.

Ezzini et al. \cite{Ezzini2021} have used a set of 39 manually defined patterns based on part-of-speech (POS) tags to detect coordination ambiguity (CA) and prepositional-phrase attachment ambiguity (PPA) \cite{Bruijn2010, AmbiguityHandbook2003}. The co-occurrence frequency of ambiguous phrase candidates is then found in the application domain of the requirement by crawling the relevant Wikipedia documents to filter the false positive results. Manually defined patterns can be fit for multiple types of ambiguity simultaneously.

In general, it is observed that Precision and Recall of semi-automated approaches are highly dependent on the manual dictionaries provided by the researchers. The precision could be improved by applying comprehensive dictionaries dictionaries not restricted to the expert's knowledge. Moreover, there is no public dataset of the natural language requirement with known smells to enable a fair comparison of different approaches.

We use NLP techniques, mainly lemmatization \cite{Jurafsky2009}, POS tagging \cite{Slav2011}, and word embedding \cite{Young2018}, to automatically detect requirement smells and measure software requirements testability, which none of the related works has used yet. Besides, we provide a publicly accessible web-based tool, ARTA, to measure natural language requirement clarity and testability.
ARTA provides the user with a text editor to modify any detected requirement smell. 
Some commercial tools use lightweight NLP techniques to manage requirement quality issues \cite{QRACorp2021, VisureSolutions2020}. However, to the best of our knowledge, ARTA is the first open-source tool dedicated to requirement quality management that is not restricted to a specific domain or template.
Other freely accessible web-based tools, such as Quality Analyser for Official Documents (QuOD) \cite{Ferrari20192} identify natural language defects in business processes. Grammarly \cite{Grammarly2021} detects issues in general English language documents that are not strictly relevant to requirements artifacts.

Our work is related to some recent papers that deal with different aspects of neural networks, such as state estimation, stability, control, fuzzing, and impulsive systems \cite{Song2023, Wei2021, Song20232, Zakeri20211, Xu2021}. These papers use neural networks to model, analyze, or control complex systems, such as reaction-diffusion neural networks, interconnected nonlinear systems, fuzzing, or impulsive systems. They also deal with the challenges or uncertainties of neural networks, such as delays, impulses, attacks, or quantization. They evaluate the performance or quality of neural networks, such as stability, robustness, or efficiency. Our work differs from these papers in that we use neural networks to generate a dictionary of smelly or context-sensitive words rather than to model or control systems. Furthermore, we evaluate the testability of requirements rather than the stability or performance of neural networks. However, our work also complements these papers in that we share some common objectives, such as understanding the complex interaction among inputs from different application domains.

\section{Methodology}\label{sec:methodology}
\noindent
We propose a novel methodology for measuring and improving the testability of natural language requirements. Our methodology consists of four main contributions: (i) a mathematical model that defines requirements testability as a function of requirements smells, size, and application domain; (ii) an automated dictionary that identifies smelly or context-sensitive words for different application domains and calculates their semantic similarity using a neural Word2Vec model; (iii) two new types of requirement smells, Polysemy and Uncertain Verbs, that affect the testability of requirements; and (iv) a public dataset of requirements testability, with annotated and validated requirements samples, smells, and testability scores. We demonstrate the correctness and applicability of our methodology through empirical studies.

This section first formalizes requirements testability, $T(R)$, as a function of requirement smells and length of the requirements as the number of sentences in the requirement definition. 
Afterward, it is described how to detect requirements smells and eventually calculate the testability. We employ a well-known goal-question-metric (GQM) approach \cite{Basili1994}, illustrated in Table \ref{table:gqm}, to build and evaluate our proposed requirements testability measurement methodology. 

The purpose is to quantify and measure the testability of the requirement artifacts (objects) defined in a natural language from the viewpoint of the requirement engineers. Quantifying requirements testability into an interpretable range of $(0,1]$ provides several benefits to requirement engineers: 
    
\begin{enumerate}[label=\roman*., noitemsep, labelwidth=!, labelindent=0pt, topsep=0pt]
\item{
To measure or estimate the cost and effort required to improve the requirement artifact before applying any validation activity, such as acceptance testing. }

\item{
To prioritize requirement artifacts based on their extent of testability and give focus on untestable parts first based on time and budget dedicated to the validation process.
}

\item{
To reveal the existing gap between the domain experts and software developers' terminologies at the beginning of the software development life cycle (SDLC) process.
}

\item{
To distinguish the requirements that support the automatic acceptance test case generation algorithms such as the one recently proposed by Fischbach et al. \cite{Fischbach2020}.
}
\end{enumerate}

In summary, our requirements testability formulation can be used as a concrete proxy to estimate the requirement debt often raised by the requirements engineering phase and help optimal decision-making from the viewpoint of project managers. 

The underlying assumption is that requirement smells negatively affect the testability of requirements. Considering a requirement artifact as a standalone objective to be met by a part of the software product, the first group of questions, in Table \ref{table:gqm}, is concerned with the quality of the requirement artifacts in terms of the number of sentences and lexicon needed to describe the requirement. The second group of questions aims at characterizing the attributes of the requirement artifact that are relevant to its testability, \emph{e.g.}, requirement smells, and finally, the third group of questions concerns evaluating the testability of requirement artifacts.

\begin{table}[!h]
    \centering
    \caption{Goal-Question-Metric (GQM) approach used to measure requirements testability.}
    \label{table:gqm}
    \resizebox{0.90\linewidth}{!}{%
        \begin{tabular}{@{}lll@{}}
            \toprule
            Goal     & \begin{tabular}[c]{@{}l@{}}Purpose\\ Issue\\ Object (products)\\ Viewpoint\end{tabular} & \begin{tabular}[c]{@{}l@{}}To measure\\ the testability of\\ the requirement artifacts\\ from the viewpoint of the requirement engineer\end{tabular}                                                                                                      \\ \midrule
            Question & Q1 Group 1                                                                              & How many sentences and words are in the requirements?                                                                                                                                                                                                 \\
            Metrics  & \begin{tabular}[c]{@{}l@{}}M1\\ M2\end{tabular}                                         & \begin{tabular}[c]{@{}l@{}}Number of sentences in the requirement\\ Number of words in the requirement\end{tabular}                                                                                                                                   \\ \midrule
            Question & Q2 Group 1                                                                              & What is the grammatical role of each word in the requirement?                                                                                                                                                                                         \\
            Metrics  & M3                                                                                      & Part Of Speech (POS) tags dictionary                                                                                                                                                                                                                  \\ \midrule
            Question & Q3 Group 2                                                                              & To which extent is the requirement smelly?                                                                                                                                                                                                            \\
            Metrics  & \begin{tabular}[c]{@{}l@{}}M4\\ M5\end{tabular}                                         & \begin{tabular}[c]{@{}l@{}}Number of smelly words in the requirements\\ Type of smells\end{tabular}                                                                                                                                                   \\ \midrule
            Question & Q4 Group 2                                                                              & How clean is a given requirement in the natural language?                                                                                                                                                                                             \\
            Metrics  & M6                                                                                      & Requirement clarity, C(R), a function of M2, M3, M4, and M5.                                                                                                                                                                                          \\ \midrule
            Question & Q5 Group 3                                                                              & \begin{tabular}[c]{@{}l@{}}What is the required test effort/ cost of automatically or manually analyzing \\ a single statement in the requirement from the viewpoint of the requirements \\ engineer?\end{tabular}                                     \\
            Metrics  & M7                                                                                      & \begin{tabular}[c]{@{}l@{}}$ \alpha \in [0,1)$, a subjective factor determined/ estimated by the requirement\\ engineer based on the inherent criticality of the requirement, the \\ application domain, and the importance of the validation process.\end{tabular} \\ \midrule
            Question & Q6 Group 3                                                                              & \begin{tabular}[c]{@{}l@{}}What is the testability value of a given requirement, defined in \\ natural languages?\end{tabular}                                                                                                                        \\
            Metrics  & M8                                                                                      & Requirement testability, T(R), a function of M1, M6, and M7.                                                                                                                                                                                          \\ \bottomrule
        \end{tabular}%
    }
\end{table}

\subsection{Formalism}\label{sec:formalism}
\noindent
Requirements testability is of great concern for acceptance testing. Requirement smells are symptoms of the poor definition of software requirements, which leads to poor testability. The more smelly words, \emph{i.e.}, words causing any kind of ambiguity in requirement definition, in a given requirement, $R$, the less testable the requirement will be:

\begin{equation}\label{eq:1}
    T(R) \propto \frac{1}{n(w_{R}^{S})}
\end{equation}
where $n(w_{R}^{S})$ is the number of smelly words in the requirement, $R$. Requirements smells and the smelly word dictionary are further described in Sections \ref{Requirement-smells} and \ref{sec:smelly-words-dictionary}.

As the length of a requirement statement increases, its testability will inversely decrease because, firstly, it is relatively more difficult to prepare test cases for a lengthy requirement definition rather than a short one. Secondly, there is a higher chance of having smells in a lengthy requirement. 

The probability of untestability grows as the number of sentences in a requirement definition increases. Good scientific writing practice suggests writing short sentences. Short sentences are the crux of good scientific writing \cite{Robertson2006}. The reason is that short sentences are easier to understand. 
Similarly, describing a requirement in one short sentence will be much easier to understand and test rather than using several sentences.
Multiple sentences of a requirement have been considered a negative factor in requirements quality by many researchers \cite{Fabbrini2000, Fabbrini2001, Alexander2002, Huertas2013}. Therefore, the testability, $T(R)$, of requirement, $R$, is also proportional to the inverse of the length of the requirement statement, $n(S_{R})$:

\begin{equation}\label{eq:2}
    T(R) \propto \frac{1}{n(S_{R})}
\end{equation}

We follow the metrics in Table \ref{table:gqm} to formulate requirements testability according to relations \ref{eq:1} and \ref{eq:2}. We begin by defining ideal definitions of natural language requirements, which form the cornerstone of our methodology.

\begin{definition}\label{def:1}
   \textbf{Clean requirement:} 
   A software requirement that does not suffer from any known requirement smell is called a clean requirement.
\end{definition}

\begin{definition}\label{def:2}
   \textbf{Fully testable requirement:} 
   A software requirement that is clean and expressed in solely one sentence is called a fully or completely testable requirement.
\end{definition}

These definitions indicate an upper bound (best-case) for two notions of \textit{requirement clarity} and \textit{requirement testability}, formulated as follows.
Considering Definition \ref{def:1}, the \textit{clarity}, $C(R)$, of the requirement, $R$, depends on the number of smelly words, $n(w_R^S)$, and the number, $t$, of different types of smells appearing in the requirement statement. 
Equation \ref{eq:3} determines requirement clarity, $C(R)$:

\begin{equation}\label{eq:3}
  C(R)= \begin{cases}
     1 &\quad\text{R is clean} \\
     1 - \left(\frac{n(w_R^S)}{n(w_{R})}\right)^{\frac{1}{t}} &\quad\text{otherwise}
  \end{cases}
\end{equation}
where $n(w_{R})$ is the total number of words in requirement $R$.

Considering Definition \ref{def:2}, the testability, $T(R)$, of a requirement, $R$, could be measured in terms of its clarity, $C(R)$, and its length, $n(S_{R})$ as follows:

\begin{equation}\label{eq:4}
T(R)=\ \frac{C(R)}{{(1+\alpha)}^{n(S_{R}) - 1}}
\end{equation}
where $\alpha \in [0,1)$, is a subjective factor, determined or estimated by the requirement engineer based on the inherent criticality of the requirement, the application domain, and the importance of the validation process. 

According to Equation \ref{eq:4}, $T(R)$ depends on both the requirement clarity, $C(R)$, and the requirement length in terms of the number of sentences in the requirement, $R$. The testability, $T(R)$, is reduced exponentially as the number of sentences, $n(S_R)$, increases based on our testability model. Moreover, the number of smelly words and the type of smells in the requirement statement negatively affect the requirement clarity, $C(R)$, which results in decreasing requirements testability. 
For example, if a requirement is ambiguous or vague, it may be challenging to design a test case verifying its fulfillment or non-fulfillment. Similarly, if a requirement is incomplete or inconsistent, it may lead to confusion or conflicts among stakeholders or developers and thus hampers the testing process \cite{Hayes2015, Beer20181, Femmer20172}.

The only hyperparameter in our testability model is $\alpha$. It determines the impact of the size, in terms of the number of the sentences, $n(S_R)$, of the requirement, $R$, on the testability, $T(R)$, of the requirement definition. 
If $\alpha = 0$ then, $T\left(R\right) = C(R)$. In this case, the number of statements in a requirement definition does not impact its testability. 
It should be noted that Definition \ref{def:2} specifies a \emph{sufficient} and not \emph{necessary} condition for a requirement to be fully testable. Indeed, a testable requirement may be expressed in more than one sentence. 
If each sentence in a requirement definition complements its former sentences, the value of $\alpha$ is $0$. However, if each sentence in a requirement definition targets a different objective, the value of $\alpha$ approaches one. In such cases, the testability, $T(R)$, is reduced exponentially as the number of sentences, $n(S_R)$, increases based on our testability model.

The value of $\alpha$ also reflects the inherent difficulty of testing a system requirement, affected by the nature of its \textit{domain}. For instance, understanding and validating a single requirement in a safety-critical system such as a self-driving car most probably need more effort than a business-critical system such as accounting. Moreover, it is relatively more challenging to test a business requirement than a simple function/non-functional requirement. If a business requirement is not defined appropriately, it does not matter how well the project is delivered; still, the business is not satisfied, and it is not right to blame the stakeholder. 
The computation of the $\alpha$ parameter is described in Section \ref{sec:guidelines-on-determining-the-alpha-factor}.

A sentence is the smallest meaningful unit of a language that can stand independently. At least one sentence is required to express a software requirement in any natural language regardless of the application domain, system criticality level, requirement type, and template. As a result, the first sentence should not impose any testing cost/effort for a given requirement. That is why our suggested formula to compute the requirement testability ignores the first sentence when computing the cost of tests. 

According to Equation \ref{eq:4}, the requirements testability is also reduced as the number of smelly words, and the type of smells in the requirement statement increases, \emph{i.e.}, requirement clarity decreases. 
More formally, our definition of requirements testability, given by $T(R)$, is based on Fuzzy logic instead of classical Binary logic, which brings a vast opportunity to use in the automated requirement analysis approaches, especially data-driven and learning-based analysis. 
The flexible definitions of equations \ref{eq:3} and \ref{eq:4} allow emerging requirement smells and issues to be considered in calculating requirements clarity and testability. Requirement smells are the subject of the next section.

\subsection{Requirements smells}\label{Requirement-smells}
\noindent
Requirements smell cause misunderstandings in the interpretation and analysis of requirements. Such misunderstandings lead to flaws in design and costly acceptance tests. The catalog of eight requirement smells, along with their detection procedure, has been proposed by Femmer et al.  \cite{Femmer20172}. 
We discovered that two requirement smells, \emph{Loopholes} and \emph{Open-ended}, are not distinguishable by their definitions \cite{Femmer20172}. Therefore, we merged them under the umbrella of \textit{Non-verifiable terms}. Moreover, we coined two new requirements smells,  named \textit{Polysemy} and \textit{Uncertain Verbs}. Table \ref{table:req-smells} summarizes the definitions of each requirement smell and its detection mechanism. 

Smells related to the grammatical aspects of requirements (S4-S7) are detected via POS tagging techniques \cite{Slav2011} in which the grammatical role of each word, \emph{e.g.}, noun, verb, adjective, adverb, or pronoun, is determined using linguistic and probabilistic models such as Hidden Markov Models \cite{Jurafsky2009}. 
Other smells (five out of nine smells) are identified using a dictionary. The \textit{lemma} of each word \cite{Jurafsky2009} within the requirement is searched in a predefined dictionary of smelly words. If the word is found in that dictionary, the requirement is tagged with the corresponding smell name. 
Uncertain Verbs could be detected based on a small dictionary of English modals. However, the other four smells (S1, S2, S3, and S9) are more challenging to detect. Section  \ref{sec:smelly-words-dictionary} proposes an automated approach to make a comprehensive dictionary of \emph{smelly words}.

\begin{table}
    \centering
    \caption{Requirement smells used to measure requirements testability}
    \label{table:req-smells}
    \resizebox{1\linewidth}{!}{%
        \begin{tabular}{llllll}
            \hline
            Smell NO. & Name                       & Definition                                                                                                                                                                                            & Example                                                                   & Detection        & Ref.          \\ \hline
            S1 & Subjective
            language        & \begin{tabular}[t]{@{}l@{}}The words for which the semantic is not defined \\ objectively. \end{tabular}                                                                                                                                            & User friendly                                                             & Dictionary       & \cite{Femmer20172}            \\
            S2 & \begin{tabular}[t]{@{}l@{}}Ambiguous\\
                adverb/adjective\end{tabular}    & \begin{tabular}[t]{@{}l@{}}Certain adverbs and adjectives that are unspecified \\ by nature.\end{tabular}                                                                                             & Almost                                                                    & Dictionary       & \cite{Femmer20172}              \\
            S3 & \begin{tabular}[t]{@{}l@{}}Non-verifiable \\
            terms  \end{tabular}   & \begin{tabular}[t]{@{}l@{}}The terms that are difficult to verify as they offer a \\ choice of possibilities or imply the fulfillment of a \\ particular but imprecisely defined extent.\end{tabular} & \begin{tabular}[t]{@{}l@{}}Sufficient, \\ as far as possible\end{tabular} & Dictionary       & \begin{tabular}[t]{@{}l@{}}This work,\\ 
                \cite{Femmer20172} \end{tabular}   \\
            S4 & Superlatives               & \begin{tabular}[t]{@{}l@{}}Superlatives refer to adjectives in requirements that \\ express a relation of the system to all other systems.\end{tabular}                                               & Highest                                                                   & POS tagging      & \cite{Femmer20172}              \\
            S5 & Comparative                & \begin{tabular}[t]{@{}l@{}}Comparatives refer to adjectives in requirements that \\ express a relation of the system to specific \\ other systems or previous situations\end{tabular}                 & More exact                                                                & POS tagging      & \cite{Femmer20172}              \\
            S6 & Negative statement         & \begin{tabular}[t]{@{}l@{}}The statements that express a specific capability not \\ to be provided by the system.\end{tabular}                                                                        & Must not sign off                                                         & POS tagging      & \cite{Femmer20172}              \\
            S7 & Vague pronouns             & \begin{tabular}[t]{@{}l@{}}The pronouns whose reference or relation is not clear \\ to the reader base on the context.\end{tabular}                                                                   & Which                                                                     & POS tagging      & \cite{Femmer20172}              \\
            S8 & Uncertain Verbs            & \begin{tabular}[t]{@{}l@{}}Verbs without reasonable certainty, less than 100\%, \\ are unclear and may cause doubt when evaluating  \\ a requirement.\end{tabular}                                     & May, can                                                                  & Small Dictionary & This work     \\
            S9 & Polysemy                   & \begin{tabular}[t]{@{}l@{}}The coexistence of many possible meanings for \\ a word or phrase base on the context.                                       \end{tabular}                                                                             & Call, return                                                              & Dictionary       & This work     \\ \hline
        \end{tabular}%
    }
\end{table}

\subsection{Smelly words dictionary}\label{sec:smelly-words-dictionary}
\noindent
Four out of nine types of smelly words, including Polysemy words, Subjective language, Ambiguous adverbs/adjectives, and Non-verifiable terms, could be looked up in a dictionary, including the smelly word and its type. 
The dictionary proposed by Femmer et al. \cite{Femmer20172} is built manually for German lexicons. However, we could automatically build a dictionary of smelly words and their smell types using word embedding techniques.
We applied a neural word embedding technique with the Continuous Bag of Words (CBOW)  architecture \cite{Mikolov2013efficient} to build the dictionary on a large corpus of Wikipedia documents and extracted 1,000 candidate words in 10 different domains of science and engineering. 
The CBOW architecture has been shown to be an effective and efficient model to find semantics relations between words in a textual corpus \cite{Mikolov2013efficient, Giatsoglou2017}.

\begin{figure}
    \centering
    \includegraphics[width=0.95\linewidth]{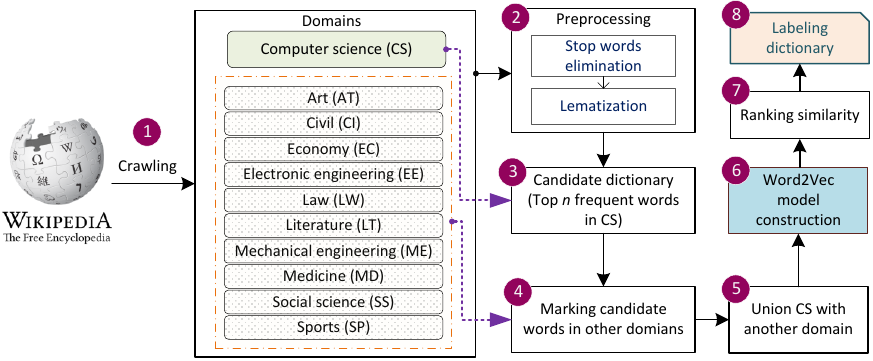}
    \caption{Dictionary building process}
    \label{fig:dictionary-building-process}
\end{figure}

The \textit{Word2Vec} algorithm, applied by the word embedding technique, provides a vector representation for each selected word. For each frequent word in computer science, we computed the cosine similarity of its vector in the computer science contexts with its vectors in the other domains. A word is considered smelly, provided that its vector similarity is lower than a given threshold. 
The selected words were then manually analyzed and labeled with the appropriate smell type. 
As a result, we could collect a dictionary of 700 smelly words and their types of smell. In this way, we could find frequent Computer Science (CS) words that appear in other contexts with different meanings. Figure \ref{fig:dictionary-building-process} shows the steps for building a dictionary. The detail of each step is discussed in the following.

First, in addition to computer science, ten Wikipedia categories, for which software systems are often generated, were selected. Each Wikipedia category is structured as a tree in which nodes are subcategories and leaves are pages \cite{Ferrari2017}. For each category, 500 subcategories were selected as different subjects, and then for each subcategory, the content of 20 pages (if any exist) with maximum length was retrieved. 
As a result, a considerable corpus of about 10K documents for each domain was selected (Step 1 in Figure \ref{fig:dictionary-building-process}). 

In Step 2, all the stop words were removed from the collected documents. Afterward, all the remaining words in the documents were lemmatized. Consequently, each word was replaced with its lemma \cite{Juergens2010}. We considered a list of 194 words, including articles, propositions, pronouns, abbreviations, numbers, and symbols, as stop words. Lemmatization is the process of transforming various inflected forms of a word (\emph{e.g.}, cats and cat's) to a single canonical word called the word's lemma (\emph{e.g.}, cat). The preprocessing results are shown in Table \ref{table:3}.

\begin{table}
    \centering
    \caption{The selected domains statistics after preprocessing}
    \label{table:3}
    \resizebox{0.60\linewidth}{!}{%
        \begin{tabular}{lllll}
            \hline
            Domain                      & Pages & Words  (W)    & Vocabulary (V) & V / W \\ \hline
            Computer science (CS)       & 9,541 & 13,575,325 & 323,500    & 0.0238           \\
            Art (AT)                    & 9,681 & 15,509,613 & 400,677    & 0.0258           \\
            Civil (CL)                  & 9,726 & 21,687,967 & 343,951    & 0.0159           \\
            Economy (EC)                & 9,856 & 23,949,093 & 409,812    & 0.0171           \\
            Electronic engineering (EE) & 9,146 & 11,537,818 & 415,568    & 0.0360           \\
            Law (LW)                    & 9,639 & 16,558,509 & 350,407    & 0.0212           \\
            Literature (LT)             & 9,960 & 17,442,460 & 551,146    & 0.0316           \\
            Mechanical engineering (ME) & 9,200 & 10,962,055 & 379,264    & 0.0346           \\
            Medicine (MD)               & 9,674 & 13,260,338 & 322,465    & 0.0243           \\
            Social science (SS)         & 9,610 & 15,520,799 & 357,477    & 0.0230           \\
            Sport (SP)                  & 9,653 & 13,438,100 & 375,577    & 0.0279           \\ \hline
        \end{tabular}%
    }
\end{table}

In the third step, we looked for and found the top 1,000 most frequent words in the preprocessed documents of CS. 
Then in Step 4, we looked for these 1,000 words in the documents collected for ten other application domains, shown in the first column of Table \ref{table:3}. All the words found in each of the documents were prefixed with an underline character. For instance, the word "return" was replaced with "\_return" in all domains apart from the CS. 
    
In Step 5, for each domain, apart from CS, in column one of Table \ref{table:3}, we created a new corpus including the documents of that domain, modified in Step 4, and the documents of CS. 
In Step 6, two word-embedding vectors were trained for each prefixed word and its corresponding word in CS. As a result, we obtained 20 vectors for each of the 1,000 frequent words in CS. Indeed, each pair of vectors belonged to a domain combined with CS. For instance, two vectors were created for the word "return" in the CS and MD domains.

In Step 7, the cosine similarities between the vector of each frequent word in CS and the other 10 domains were computed. The computed cosine similarities for each word were averaged. Afterward, we sorted the similarities ascendingly. The cosine similarity between two vectors $\vec{v_{1}}$ and $\vec{v_{2}}$ is given by Equation \ref{eq:cosine_similarity}:

\begin{equation}\label{eq:cosine_similarity}
    Sim_{cos}(v_1, v_2) = \frac{\vec{v_{1}} . \vec{v_{2}}}{|\vec{v_{1}}| \times |\vec{v_{2}|}}
\end{equation}

In Step 8, starting with the highest-ranked word, we determine its type of smell from within the four predetermined smells of Polysemy words, Subjective language, Ambiguous adverbs/adjectives, and Non-verifiable terms. 
As a result, discussed in Section \ref{sec:smelly-words-dictionary-analysis}, 700 smelly words were selected, and together with their type, they established our dictionary of smelly words.

Algorithm \ref{alg:dictionary-buiding} shows the pseudo-code of the dictionary-building process. It receives as input a list of application domain names, \texttt{domains}, number of frequent words, \texttt{n}, number of subcategories to crawl, \texttt{sub\_cats}, and number of pages to retrieve for each domain, \texttt{pages}. 
The remaining three inputs are Word2Vec algorithm hyperparameters, including the vector's dimension used to represent each word in the corpus, \texttt{dim}, the minimum occurrence of the word to be considered in the learning process, \texttt{min}, and the number of neighbors of each word, \texttt{window}. The algorithm returns a dictionary of smelly words as its output.

\begin{figure}
    \begin{center}
        \begin{minipage}{0.95\linewidth}
            \vspace{-1mm}
            \begin{algorithm}[H]
                 \small
                \onehalfspacing
                \caption{BuildDictionary} \label{alg:dictionary-buiding}
                \DontPrintSemicolon
                \setcounter{AlgoLine}{0}
                \LinesNumbered
                \SetKwFunction{Document}{Document}
                \SetKwFunction{DocumentList}{DocumentList}
                \SetKwFunction{List}{List}
                \SetKwFunction{Wikipedia}{Wikipedia}
                \SetKwFunction{POSTagger}{POSTagger}
                \SetKwFunction{FrequentWord}{FrequentWord}
                \SetKwFunction{WordToVec}{Word2Vec}
                \SetKwFunction{WordToVecList}{Word2VecList}
               \SetKwFunction{Merge}{Merge}
               \SetKwFunction{CosineSimilarity}{CosineSimilarity}
               \SetKwFunction{Dictionary}{Dictionary}
                \SetKwFunction{len}{len}
                \SetKwInput{KwData}{Input}
                \SetKwInput{KwResult}{Output}
                \KwData{List  \texttt{Domain}, int \texttt{n}, \texttt{sub\_cats}, \texttt{pages}, \texttt{dim}, \texttt{min}, \texttt{window}}
                \KwResult{Dictionary \texttt{ranked\_words}}
                \BlankLine
                \tcc{{\color{Green}1. Add Computer Science as the first (No. 0) element to the list of domains defined by the user}}
               \texttt{domains}.$insert$(\textit{“computer\_science”}, $0$)\;
               
               \tcc{{\color{Green}2. Create a corpus of cleaned (lemmatized, and without stop words) documents.}}
             \DocumentList \texttt{docs} [ ] $\leftarrow$ \DocumentList[\len(\texttt{domains})]
             
              \ForEach{ \texttt{domain} in \texttt{domains}} {
                 \Document \texttt{doc} $\leftarrow$ \Wikipedia.$crawl$(\texttt{domains}, \texttt{sub\_cats}, \texttt{pages})\;
                 
                 \texttt{doc}.$remove\_stop\_words$()\;
                 
                 \texttt{doc}.$lemmatize$()\;
                 
                 \texttt{docs}.$append$(\texttt{doc})\;
             }
                
                \tcc{{\color{Green}3. Create a Word2Vec model for the words in each domain }}
                \List \texttt{frequent\_words} $\leftarrow$ doc[$0$].$get\_frequent\_words$(\texttt{n})\;
                
                \WordToVecList \texttt{models} [ ] $\leftarrow$ \WordToVecList[\len(\texttt{domains})]\;
                
                \For{ $i \leftarrow 1$ \KwTo \len(\texttt{docs}) $-1$}{
                    \ForEach{\texttt{word} in \texttt{docs}[$i$].$words$}{
                        \If{\texttt{word} $\in$ \texttt{frequent\_words}}{
                                \texttt{word} $ \leftarrow $ "\_" + \texttt{word}\;
                          } 
                     }
                \WordToVec model $ \leftarrow $ \WordToVec(\Merge(\texttt{doc}[$0$], \texttt{doc}[$i$]), \texttt{dim}, \texttt{min}, \texttt{window})\;
                
                 \texttt{models}.$append$(\texttt{model})\;
            }
        
               \tcc{{\color{Green}4. Create and fill smelly word dictionary}}
                \Dictionary \texttt{ranked\_words} = \Dictionary()\;
                
                \ForEach {\texttt{word} in \texttt{frequent\_words}}{
                \texttt{sum} $\leftarrow$ $0$ \;
                
                 \ForEach{\texttt{model} in \texttt{models}}{
                    \texttt{vec1} $\leftarrow$ \texttt{model}.$vectors$(\texttt{word})\;
                    
                    \texttt{vec2} $\leftarrow$ \texttt{model}.$vectors$(“\_” + \texttt{word})\;
                    
                    \texttt{sim} $\leftarrow$ \CosineSimilarity(\texttt{vec1}, \texttt{vec2})\;
                    
                    \texttt{sum} $\leftarrow$ \texttt{sum} + \texttt{sim};
                }
                \texttt{sim\_avg} $\leftarrow$  \texttt{sum} $/$ \len(\texttt{domains})\;
                
                \texttt{ranked\_words}.$update$(\texttt{word}, \texttt{sim\_avg})\;
                
            }
        
                \tcc{{\color{Green} 5. Sort the smelly word dictionary and return the result}}
                \texttt{ranked\_words}.$values$().$sort\_ascending$()\;
                
               \textbf{return} \texttt{ranked\_words}\;

            \end{algorithm} 
        \end{minipage}
    \end{center}
\end{figure}

\subsection{Cost of multiple sentences}\label{sec:guidelines-on-determining-the-alpha-factor}
\noindent
In the proposed requirements testability model, basically, we compute the clarity of a requirement as a factor directly affecting its testability. However, if a requirement has more than one sentence, besides the requirement clarity, the impact of the sentences on each other and the effort required to test the requirement should be considered. We use the $\alpha$ parameter to compute the extra cost/effort for multi-sentence requirement definitions.

\begin{figure}
    \centering
    \includegraphics[width=0.75\linewidth]{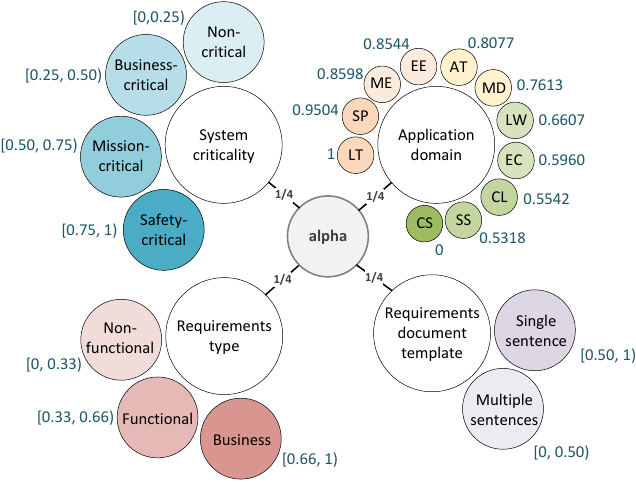}
    \caption{Guidelines to determine the alpha as the base cost of extra sentences in a requirement}
    \label{fig:extra-sentence-cost}
\end{figure}

We pinpoint four different aspects of the parameter alpha, summarized in Figure \ref{fig:extra-sentence-cost}, as mentioned in Section \ref{sec:formalism}. These factors include application domain, system criticality, requirements type, and requirements document template. Each of these factors participates equally in determining the value of alpha. Equation \ref{eq:alpha-computation} is used to compute alpha:
    
    \begin{equation}\label{eq:alpha-computation}
       \alpha = \frac{1}{4} (diss_{normalized}(D)  + CriticalityLevel +RequirementsType+DocumentTemplate )
    \end{equation}

The dissimilarities of various domains with computer science are measured by applying a word embedding technique described in Section \ref{sec:smelly-words-dictionary}. The range of values for the other three parameters is determined by equally partitioning the interval $[0,1)$ into the number of possible options for each parameter. After selecting the interval, the exact value for each of these parameters is computed by applying a severity policy. 
Such policies are commonly applied to compute code smells \cite{Fontana2015}. Two kinds of \textit{hardened} and \textit{softened} policies may be applied. The hardened policy considers the \textit{maximum} allowed value for each parameter option. In contrast, the softened policy uses the \textit{minimum} possible value in the range of the possible values for the option. 

The \emph{application domain} parameter measures the distance between the computer science and the requirement's domain. For instance, the civil engineering domain has fewer polysemy words in common with computer science than electrical engineering. The distance between computer science and the other domains depends on the vector similarity of the words most frequently used in computer science with the same words used in the other domains. We applied Algorithm \ref{alg:dictionary-buiding} to compute vector similarities, described in Section \ref{sec:smelly-words-dictionary}. The dissimilarities between the computer science domain, $CS$, and each application domain, $D$, are computed as follows: 
    
  \begin{equation}\label{eq:domain_dissim}
    dissim\ (D)\ =\ \left(1-{avg}_{sim}(D,\ CS)\right)\times\ \frac{V(D)}{W(D)}
\end{equation}

In Equation \ref{eq:domain_dissim}, $dissim\ (D)$ indicates the dissimilarity between the application domain, $D$, and computer science; $avg_{sim}(D, CS)$ represents the average cosine similarity between the most frequent words in computer science and domain $D$; $V(D)$ is the vocabulary size of the collected documents for $D$; $W(D)$ is the number of words in the collected documents for the domain $D$.
The dissimilarity between an application domain, $D$, and the CS domain can be measured as the average dissimilarity of their common frequent words in the two domains. The average dissimilarity can be defined as '$1- \text{average similarity}$'. 

In Equation \ref{eq:domain_dissim}, the average dissimilarity is also multiplied by the ratio of vocabulary size, $|V(D)|$, to all words, $|W(D)|$, in a collected corpus for the domain, $D$, to consider a difficulty factor of words in the domain. Therefore, whenever the average dissimilarity for two application domains with the CS domain is equal, the domain whose vocabulary is larger or whose total of words is smaller is considered more dissimilar to CS. The reason is that a domain with many distinct words (large vocabulary size) or a domain for which fewer documents can be found is more difficult to understand than a domain with a small vocabulary or many documents.
To make this idea more concrete, let's consider two hypothetical domains: $A$ and $B$. Suppose that both domains have the same average dissimilarity with CS, but domain $A$ has a vocabulary size of 1,000 words and a total of 10,000 words in its corpus, while domain $B$ has a vocabulary size of 500 words and a total of 20,000 words in its corpus. According to Equation  \ref{eq:domain_dissim}, domain $A$ would be more dissimilar to CS than domain $B$ because it has a larger ratio of vocabulary size to total words: $\frac{|V(A)|}{|W(A)|}  = 0.1$, while $\frac{|V(B)|}{|W(B)|}  = 0.025$. This means that domain $A$ has more unique or rare words than $B$ that are not common in CS or that domain $A$ has fewer available documents to learn from than domain $B$. Therefore, domain $A$ is more difficult to understand than domain $B$ from the perspective of CS.

Table \ref{table:domain-dissim} shows the $dissim\ (D)$ value computed by Equation \ref{eq:domain_dissim} for each domain $D$. The last row represents the normalized value of d$issim\ (D)$, computed by Equation \ref{eq:domain_dissim_norm}.

\begin{equation}\label{eq:domain_dissim_norm}
 {dissim\ }_{normalized}(D)\ =\ \frac{dissim\ (D)\ -\ \min\limits_{d\in Domains}{dissim\ (d)}\ }{\max\limits_{d\in Domains}{dissim\ (d)}-\min\limits_{d\in Domains}{dissim\ (d)}}
\end{equation}
where $Domains=\{CS,\ SS,\ CL,\ EC,\ LW,\ MD,\ AT,\ EE,\ ME,\ SP,\ LT\}$.
    
Equation \ref{eq:domain_dissim_norm} normalizes dissimilarities to the range [0, 1]. According to the last row of Table \ref{table:domain-dissim}, the dis-similarities of the ten application domains with computer science in the descending order of their magnitude are Literature, Sport, Mechanical Engineering, Electrical Engineering, Art, Medicine, Law, Economy, Civil, Social Science, and Computer Science.
It is observed that Literature, Sport, and Mechanical Engineering are more difficult to understand than Social Science, Civil Engineering, and Economic domains.
If the requirements belong to more than one domain, then ${dissim}_{normalized}(D)$ shall be computed by averaging all the relevant domains.
    
    \begin{table}
        \centering
         \caption{The application domain dissimilarities, computed for ten different domains and used in our study}
        \label{table:domain-dissim}
        \resizebox{0.90\linewidth}{!}{%
            \begin{tabular}{lllllllllll} 
                \hline
                \rowcolor[rgb]{0.949,0.949,0.949} Domain (D) & SS     & LW     & EC     & CL     & AT     & LT     & EE     & ME     & SP     & MD      \\ 
                \hline
                ${avg}_{sim}(D,\ CS)$ & 0.6288 & 0.4997 & 0.4405 & 0.4404 & 0.4974 & 0.4920 & 0.6190 & 0.6011 & 0.4531 & 0.4970  \\
                $1-{avg}_{sim}(D,\ CS)$ & 0.3712 & 0.5003 & 0.5595 & 0.5596 & 0.5026 & 0.5080 & 0.3810 & 0.3989 & 0.5469 & 0.5030  \\
                $dissim\ (D)$ & 0.0085 & 0.0106 & 0.0096 & 0.0089 & 0.0130 & 0.0161 & 0.0137 & 0.0138 & 0.0153 & 0.0122  \\
                \rowcolor[rgb]{1,0.949,0.8}               ${dissim}_{normalized}(D)$     & 0.5318 & 0.6607 & 0.5960 & 0.5542 & 0.8077 & 1      & 0.8544 & 0.8598 & 0.9504 & 0.7613  \\
                \hline
            \end{tabular}
        }
    \end{table}

\emph{System criticality}, shown in Figure \ref{fig:extra-sentence-cost}, is a second parameter used in the computation of alpha. 
Regarding system criticality, we consider four levels of non-critical, business-critical, mission-critical, and safety-critical systems
 \cite{Dubrova2013}. 
As discussed in the manuscript, understanding and validating a single requirement in a safety-critical system, such as a self-driving car, presumably, needs more effort than a business-critical system, such as accounting. In this respect, as shown in Figure \ref{fig:extra-sentence-cost}, we assign a criticality value in the intervals of $[0, 0.25)$, $[0.25, 0.5)$, $[0.5, 0.75)$, and finally $[0.75, 1]$, to the non-critical, business-critical, mission-critical and finally safety-critical systems, respectively.
    
\emph{Requirements types}, typically, are non-functional, functional, and business \cite{Sommerville2016, Goldsmith2004}. Testing a business requirement is relatively more difficult than a simple, functional, or non-functional requirement. If a business requirement is not defined appropriately, it does not matter how well the project is delivered; the business still will not be satisfied, and it is not right to blame the stakeholder. We equally divided the interval of $[0, 1)$ into three parts. Each part reveals the range for extra sentences' cost according to the importance of the requirement types, shown in Figure \ref{fig:extra-sentence-cost}.
It should be noted that extra sentences make the non-functional requirements, which are inherently subjective, more testable due to better descriptions of these requirements. 

Finally, the \emph{requirements document template} parameter determines whether the requirements definition follows a specific predefined format, \emph{e.g.}, "\textit{System shall do X}." We consider a relatively higher cost for violation of or non-conformance to the chosen template than the cost of having no templates in cases of multiple sentences requirements \cite{Arora2015}.

\subsection{Requirements testability measurement algorithm}
\noindent
Our smell detection and requirements testability measurement algorithm is shown in Algorithm \ref{alg:2}. It receives as input a requirement, \texttt{R}, a smelly words dictionary, \texttt{SD}, described in Section \ref{sec:smelly-words-dictionary}; the POS dictionary, \texttt{PD}, containing POS tags and their corresponding smells, shown in Table \ref{table:req-smells}; and the cost, \texttt{$\alpha$}, of testing each sentence in the requirement, \texttt{R}. It returns the testability value, \texttt{T}, and a list of smelly words, \texttt{LS}, and their smell types observed in the given requirement, \texttt{R}. 

At first, the algorithm splits the input requirement, \texttt{R}, into a list of sentences, and the POS tag for each word in each of the sentences is determined using a \texttt{POSTagger} function. Then, for each word in the \texttt{pos\_tags} list, the grammatical role of the word is searched in the POS dictionary, \texttt{PD}, to find possible smells with the POS tagging strategy. If the word does not have any smell, it is searched in the smelly words dictionary, \texttt{SD}, to find dictionary-based smells. In both conditions, if a smell is detected, the smelly word and its type of smell will be added to \texttt{LS}. Finally, the clarity score and testability for the requirement, \texttt{R}, are computed according to Equations \ref{eq:3} and \ref{eq:4}, respectively.

\begin{figure}[!h]
    \begin{center}
  \begin{minipage}{0.95\linewidth}
       \vspace{-1mm}
        \scriptsize 
        \begin{algorithm}[H]
            \small
            \onehalfspacing
            \caption{FindRequirmentSmellsAndTestability} \label{alg:2}
            \DontPrintSemicolon
            \setcounter{AlgoLine}{0}
            \LinesNumbered
            \SetKwFunction{Split}{Split}
            \SetKwFunction{POSTagger}{POSTagger}
            \SetKwFunction{power}{power}
            \SetKwFunction{len}{len}
            \SetKwFunction{set}{set}
            \SetKwInput{KwData}{Input}
            \SetKwInput{KwResult}{Output}
            \KwData{Requirement  \texttt{R}, SmellyWordsDictionary \texttt{SD}, 
                POSDictionary \texttt{PD},  float \texttt{$\alpha$}
            }
            \KwResult{Testability \texttt{T}, ListOfSmells \texttt{LS}}
            \BlankLine
            float \texttt{clarity}, \texttt{T};
            
            list \texttt{LS};
            
            \texttt{sentences}  $\gets$ \Split(\texttt{R.Text}) \tcc*[f]{{\color{Green}List of sentences in the requirement text}}\;
            
            \texttt{pos\_tags}  $\gets$ \POSTagger(\texttt{sentences}) \tcc*[f]{{\color{Green}List of tokens in the requirement sentences, containing (word, tag) tuples}}\;
            
            \ForEach{\texttt{token} in \texttt{pos\_tags}} {
                    \uIf(\tcc*[f]{{\color{Green}If the word tag is found in \texttt{PD}, \emph{i.e.}, its grammatical role, is reckoned as a smell}}){\texttt{PD}[\texttt{token.Key}] is not $null$}{\texttt{LS.add}(\texttt{token.Word}, \texttt{PD}[\texttt{token.Tag}]) \;
                     }
                     \ElseIf(\tcc*[f]{{\color{Green}If word found in SD, \emph{i.e.} it is detected by the automated dictionary}}){\texttt{SD}[\texttt{token.Word}] is not $null$}{\texttt{LS.add}(\texttt{token.Word}, \texttt{SD}[\texttt{token.Word}]) \;
                 }
             } 
           
            \eIf(\tcc*[f]{{\color{Green}If no smell has been found in the requirement}}){\len(\texttt{LS}) == 0}{
                    \texttt{clarity} $\gets$ $1$ \;
                    }
                    {
                           float \texttt{smellness} $\gets$  \len(\texttt{LS}) / \len(\texttt{pos\_tags})\;
                           
               \texttt{clarity} $\gets$ 1 - \power(\texttt{smellness},  1 / \len(\set(\texttt{LS}[$:, 1$])) )\;
                    }
          
          \texttt{T} $\gets$   \texttt{clarity} / \power($1+\alpha$, \len(\texttt{sentences})-1)\;
          
            \textbf{return} (\texttt{T}, \texttt{LS})\;

        \end{algorithm} 
   \end{minipage}
\end{center}
\end{figure}

\subsection{Design and implementation}
\begin{figure}[!h]
    \centering
    \includegraphics[width=0.55\linewidth]{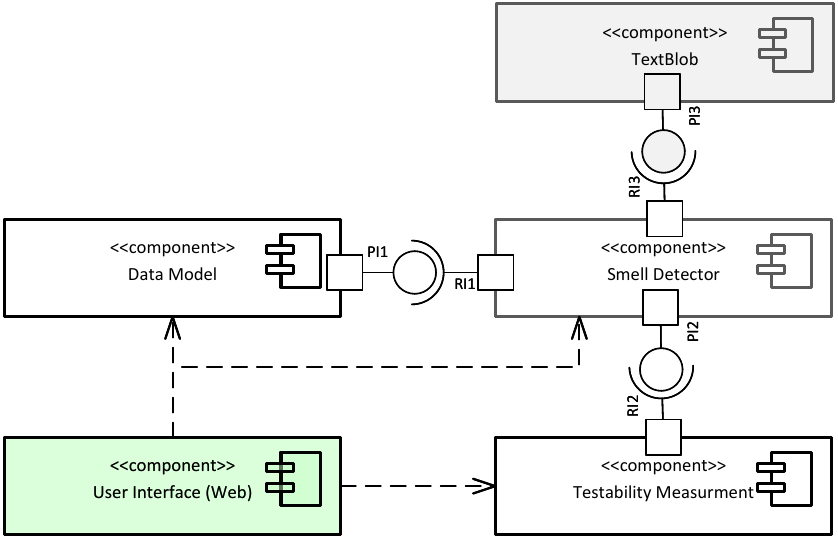}
    \caption{ARTA component diagram}
    \label{fig:arta-component-diagram}
\end{figure}

\noindent
Our proposed method for requirements testability consists of four components, illustrated in Figure \ref{fig:arta-component-diagram}. The Data Model component contains project identities, their requirements, and smells labeled by experts. The Smell Detector component detects the smells in a requirement definition.
If a smell is found that does not exist in our smell list, then we can easily add its detection strategy to this component.
The Testability Measurement component calculates the testability value for each requirement in the project, and the fourth component, User Interface, provides data visualization and create-read-update-delete (CRUD) operations for the projects, requirements, and smells. 

The proposed system is implemented in Python. We use TextBlob \cite{TextBlob2020}, a library for typical natural language processing tasks, for requirement preprocessing and POS tagging. Word embedding learning is performed by Gensim \cite{Gensim2020}, which is a Python topic modeling and semantic analysis library. Finally, our web-based user interfaces are implemented by Django \cite{Django2020}. Django is a high-level Python web framework that encourages rapid development. The source code of ARTA is available on the public GitHub repository \textit{\href{https://github.com/m-zakeri/ARTA}{https://github.com/m-zakeri/ARTA}}.

\subsubsection{ARTA web interface}
\begin{figure}[!h]
    \centering
\begin{subfigure}
    {0.49\linewidth}
    \centering
    \frame{\includegraphics[width=0.99\linewidth]{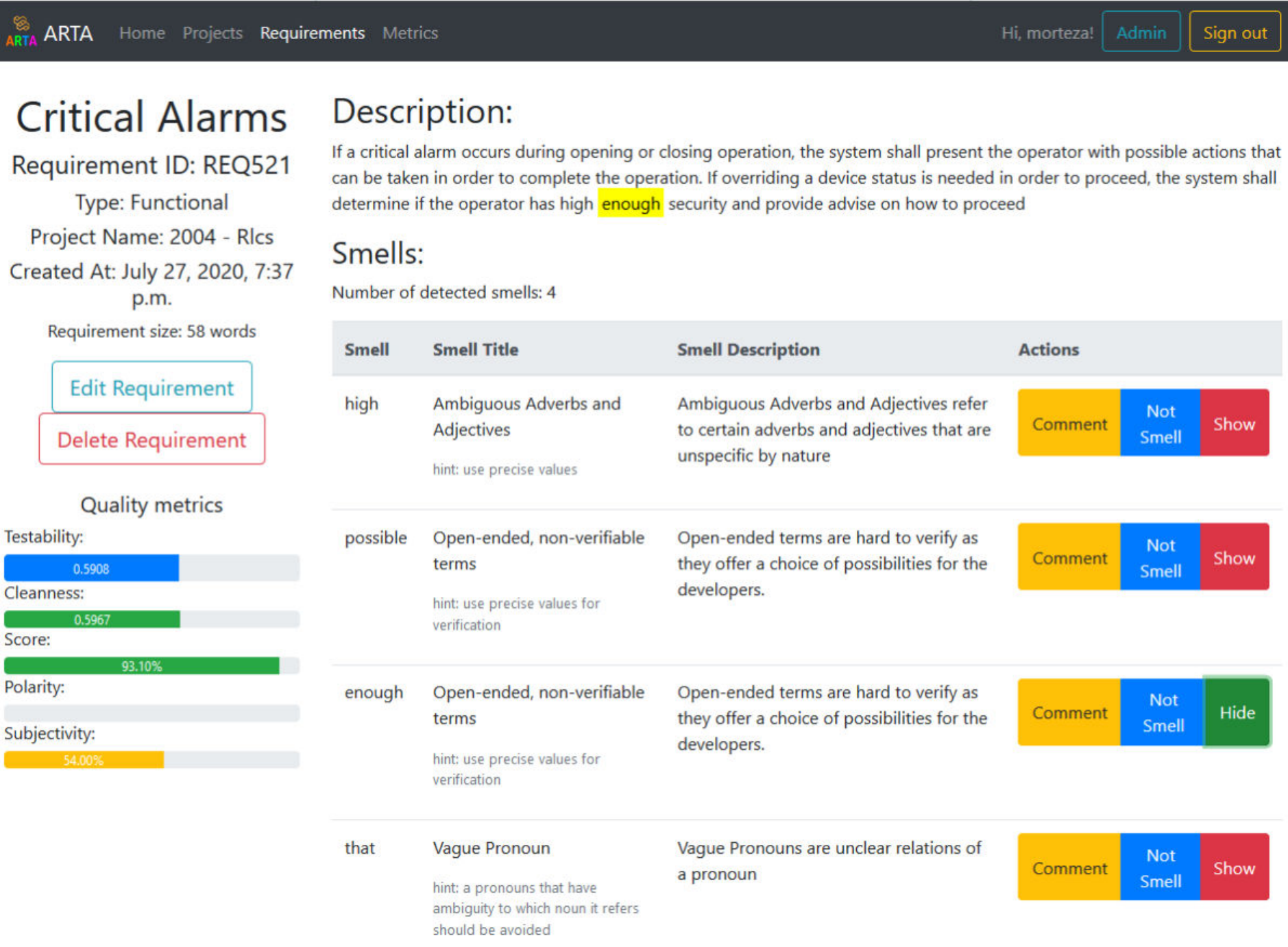}}
    \caption{Requirement analyzer module}
    \label{fig:arta-web-interface-module-i}
\end{subfigure}
    \begin{subfigure}
        {0.49\linewidth}
        \centering
        \frame{\includegraphics[width=0.99\linewidth]{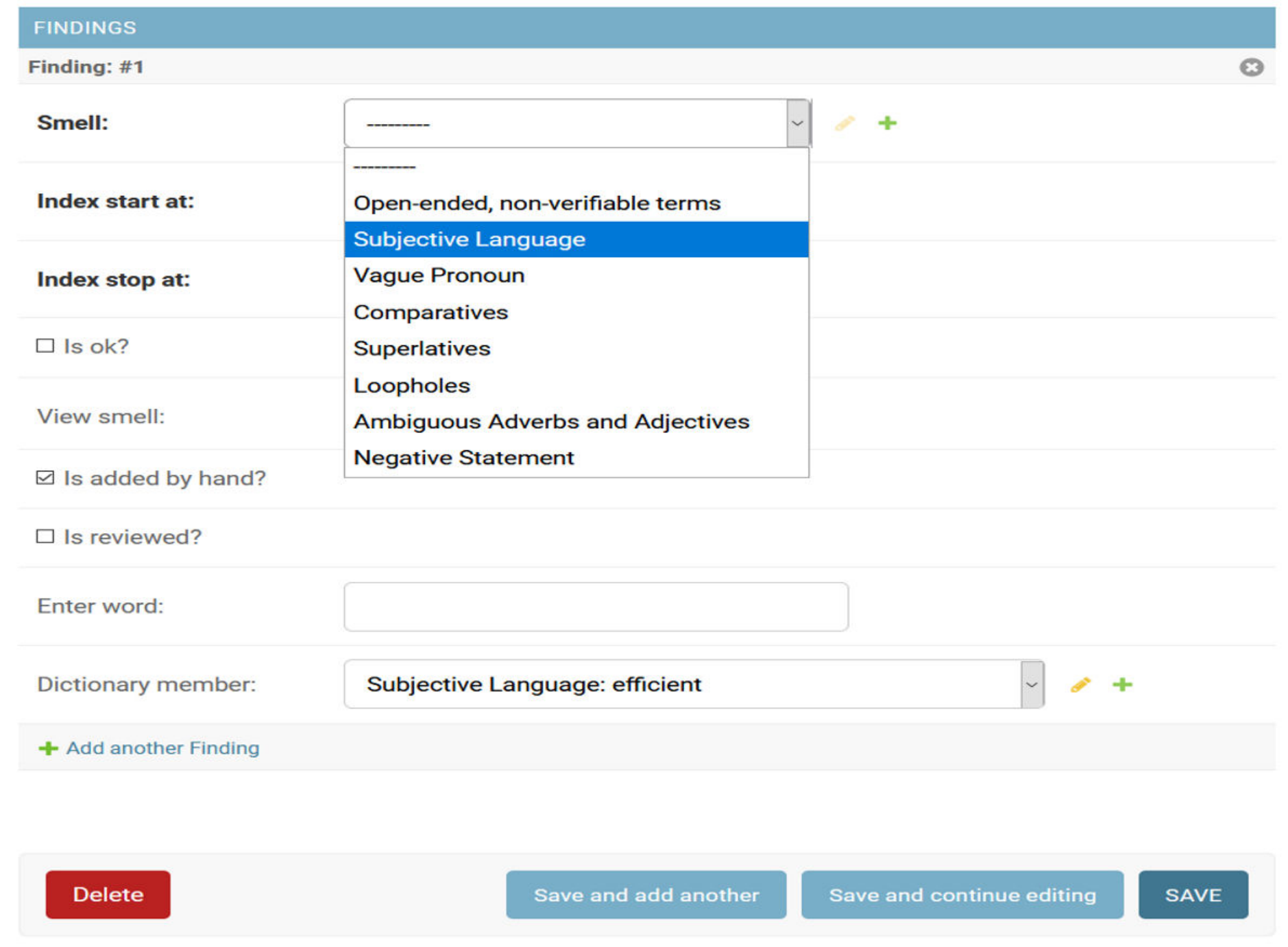}}
        \caption{Requirement smells labeling module}
        \label{fig:arta-web-interface-module-ii}
    \end{subfigure}
 \caption{ARTA web interface}
\label{fig:3}
\end{figure}

\noindent
As described in Section \ref{sec:introduction}, our system has been implemented as a standalone web application. Once a requirement is added to a project, the system automatically detects requirement smells and computes different quality metrics, including requirements testability and clarity, shown in Figure \ref{fig:arta-web-interface-module-i}. ARTA also highlights smells detected in a requirement definition and notifies the requirement engineer to refactor the requirement, considering its smell type. In addition to automatically detecting the smells, our Automatic Requirements Testability Analyzer, ARTA, allows specifying the requirements smells manually.

The second important module in ARTA is the Manual Labeling module. It helps researchers specify smells found in a requirement definition and provides a review flag that the reviewers can use to ensure the correctness of the smell type assigned to the requirement.
Figure \ref{fig:arta-web-interface-module-ii} shows the ARTA manual smell labeling module. For the time being, since a large dataset of textual requirements with known smells is not available, machine learning-based techniques for smell detection are not applicable. 
The labeling module provided by ARTA accelerates the process of creating the desired dataset to develop and evaluate smell detection algorithms.

\section{Evaluation}\label{sec:evaluation}
\noindent This section reports the setup and the result of our proposed methodology evaluation.

\subsection{Experimental setup}

\subsubsection{Research questions}
 \noindent 
We design our experiments to answer the following research questions about the requirement smells and requirements testability:
 
\begin{itemize}[noitemsep, labelwidth=!, labelindent=0pt, topsep=0pt]
    \item{\textbf{RQ\textsubscript{1} (Requirement smells prevalence)}
        What is the prevalence of requirement smells in real-world software projects? Which smells are more frequent than others? Is there a relation between requirement smells and the size and number of requirements in projects?
    }

    \item{\textbf{RQ\textsubscript{2} (Smelly words dictionary)}
       Which frequently used words in computer science are context-sensitive and thereby cause ambiguity?
    }

    \item{\textbf{RQ\textsubscript{3} (Requirement smells detection performance)}
        How effective is the automated dictionary mechanism in spotting smells in software requirements?} Which smells are more difficult to detect?

    \item{\textbf{RQ\textsubscript{4} (Requirements testability measurement performance)}
        How effective is the proposed method in measuring the testability of the requirements? Which smells affect requirements testability more than the others?
    }
\end{itemize}

\subsubsection{Dataset}\label{sec:case-studies}
\noindent
We evaluated our proposed methodology on a set of real-world software requirements obtained from the PURE dataset \cite{Ferrari20170}. It consists of requirements documentation of $79$ industrial and academic projects. 
In this repository, the description of all the included projects and their requirements is available in PDF, DOC, HTML, and XML formats. The collector of the PURE dataset has already extracted the requirement of 18 projects into XML files, and we selected these projects as the initial set of our case studies. 

Manual determination of the requirements smell is time-consuming because each requirement should be studied and analyzed carefully several times by at least two experts to achieve reliable results. For the scope of this paper, we chose six real-life projects with different domains and a different number of requirements. We manually extracted the requirements definitions of these projects into a Microsoft Excel file and labeled each requirement with its smell types. Table \ref{table:selected-projects} shows the projects and the number of requirements for each project. In total, 985 requirements were obtained from six projects. It is worth noting that our primary dataset consists of 5,000 requirements from 15 projects. Each of the requirements was labeled by four groups of IUST software engineering students. However, based on the consensus, we chose a subset of requirements with the most reliable labels after the second and third revisions.

\begin{table}
    \centering
    \caption{Selected projects and their requirements}
    \label{table:selected-projects}
    \resizebox{0.80\linewidth}{!}{%
        \begin{tabular}{lllll}
            \hline
            Project    & Topic                            & Scope / Domain                                                                    & \begin{tabular}[c]{@{}l@{}}Number of\\ requirements\end{tabular} & \begin{tabular}[c]{@{}l@{}}Number of\\ words\end{tabular} \\ \hline
            EIRENE     & Digital radio standard for railway                & Industry / EE                                                  & 564                                                              & 14,707                                                     \\
            ERTMS/ETCS & Train control system                  & Industry / EE, ME                                                              & 199                                                              & 3,566                                                      \\
            CCTNS      & \begin{tabular}[c]{@{}l@{}}Crime investigation \\ management system\end{tabular}           & Industry  /  LW        & 115                                                              & 3,635                                                      \\
            Gamma-J    & Web store             & Academic / EC, CS                                                                               & 51                                                               & 683                                                        \\
            KeePass    & Password management system              & Industry  / CS                                                            & 32                                                               & 456                                                        \\
            Peering    & \begin{tabular}[c]{@{}l@{}}Internetworking of content delivery\\ network through Peering\end{tabular} & Academic / CS & 24                                                               & 195                                                        \\
            \multicolumn{3}{l}{\textbf{Sum}}                                                                                            & \textbf{985}                                                              & \textbf{23,242}                                                     \\ \hline
        \end{tabular}%
    }
\end{table}

\subsubsection{Dataset creation procedure}\label{sec:requirement-dataset}
\noindent
We employed a cross-labeling process in which each requirement was labeled by one student and then reviewed by two other students. Four groups of IUST postgraduate (2\textsuperscript{nd}-semester M.Sc.) students in the field of software engineering were chosen during the Advanced Software Engineering (ASE) course presented by the corresponding author of this paper. At least one of the students in each group had relevant industrial experiences. 
All the postgraduate students who participated in the labeling process were asked to carefully study the ISO/IEC/IEEE 29148:2011 \cite{ISO/IEC/IEEE20112}, titled \textit{“Systems and software engineering --- Life cycle processes --- Requirements engineering”} to get insights into the standard definitions of requirement smells and dictionaries. Students were also asked to look at similar standards \cite{Schneider2013} and study recent papers in the field of requirements engineering \cite{Femmer20172, Ferrari2019, Zhao2021}. Students were then trained for two 1.5-hour sessions. The training was about the concept of requirement smells and smell detection mechanisms and the process of labeling. 

\begin{figure}
    \centering
    \includegraphics[width=0.45\linewidth]{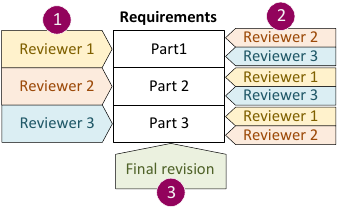}
    \caption{Cross-labeling process}
    \label{fig:cross-labeling-process}
\end{figure}

The selected projects were randomly assigned to each group of students, and we asked each group to study the project documents to get ideas about the scope and domain of the projects. Figure \ref{fig:cross-labeling-process} shows the cross-labeling process used in our dataset creation phase. The requirements of all projects in each group were divided into three parts. Each part was labeled by a postgraduate in software engineering. 
Each student also reviewed the result of labeling by two other teammates. In cases of disagreement, the results were resolved by discussion and interlocution. Using this cross-labeling mechanism, we could ensure that smells were determined carefully, and that our results were as reliable as possible. Finally, all requirements were reviewed by a Ph.D. student in software engineering (first author of the paper). 
In cases where it is complicated to decide if the requirement has a specific smell, we simply removed the requirement from our final dataset due to the lack of expert knowledge or different labels.

The overall labeling process took about three months. Unfortunately, due to different students' skills, different nature of projects, and lack of time, we could not confirm the label of all requirements in all the projects. We discarded projects in which the reviewers could not evaluate more than 10\% of their requirements. In addition, one group could not complete the labeling process.

Finally, we ended up with 985 requirements that both the experts in software industries and our teams agreed on their type of smells. We observed an average Fleiss’ Kappa of 0.9 for the six confirmed projects, which shows almost perfect agreement and confirms our dataset set is based on a reliable consensus. This is the first public requirement dataset labeled by the known smells on its own.

\begin{figure}
    \centering
    \includegraphics[width=1.0\linewidth]{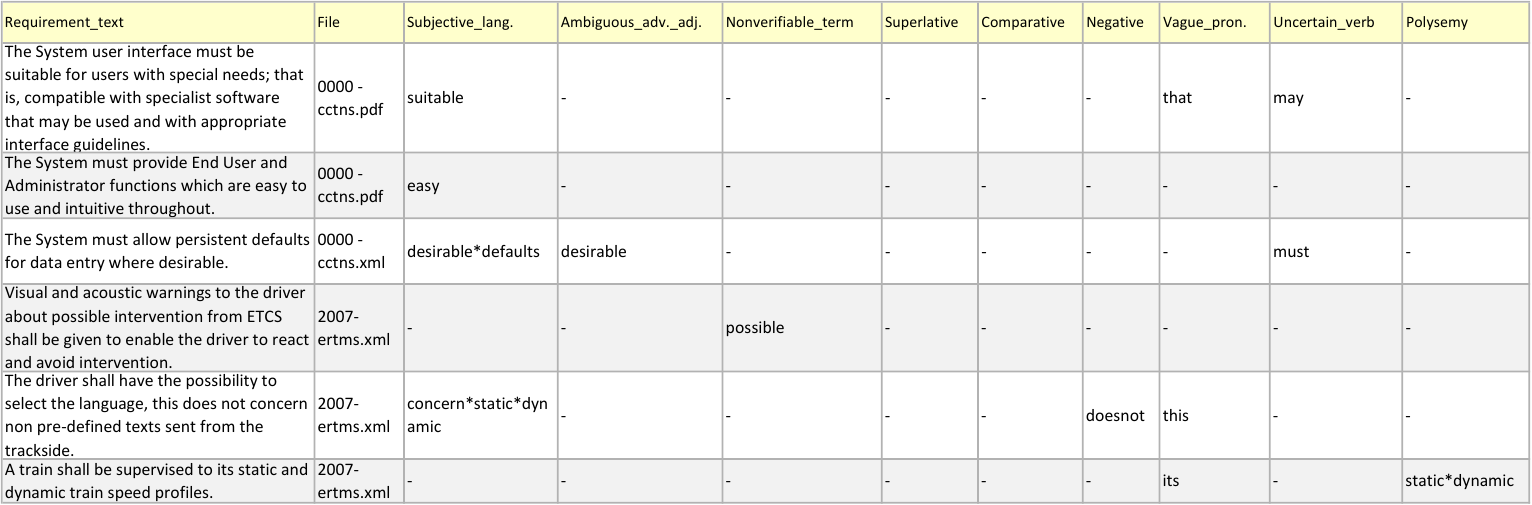}
    \caption{Requirements dataset schema with samples}
    \label{fig:dataset-schema}
\end{figure}

As described in Section \ref{sec:case-studies}, all the requirement definitions were labeled and stored in an Excel file. This Excel file contains 11 columns, including the requirement text, project file, and one column for each smell. Figure \ref{fig:dataset-schema} shows the Excel datasheet schema and the five sample requirements.
The smelly word for each requirement is written in the cell $(i,\ j)$, where $i$ is the row requirement, and $j$ is the column of smell. Therefore, our dataset not only points out the smell but also locates the position of the smell within the text of the requirements. If a requirement has more than one word with the same smell type, \emph{e.g.}, two subjective terms, these words are separated by an asterisk (*) symbol in the cell $(i, j)$. Empty cells are marked by a '\texttt{-}' character.
This template facilitates quick and easy retrieval of requirements to evaluate requirement smells and testability. 

\subsubsection{Analysis procedure} 
\noindent
Algorithm \ref{alg:dictionary-buiding} was applied to create our smelly word dictionary using application domains listed in Table \ref{table:3}. We set \texttt{n}$=1,000$, \texttt{sub\_cats}$=500$,  \texttt{pages}$=20$,  \texttt{dim}$=50$,  \texttt{min}$=5$,  \texttt{window}$=10$, for our experiments.
 Afterward, we ran Algorithm \ref{alg:2} to find smells in reach requirement and compute testability. The results of both algorithms are saved in Excel files. Our experiments are performed by analyzing the data of these Excel files.

We used the Google Colaboratory platform \cite{Colab2020} to crawl Wikipedia and build Word2Vec models, which is a free cloud-based computational resource. All the experiments were performed on a Windows 10 (x64) machine with a 2.6 GHz Intel\textregistered~Core\texttrademark~i7 6700HQ CPU and 16GB of RAM. 

Neither the implementation nor the dataset of Smella \cite{Femmer20172} is publicly available. Therefore, we reimplemented Smella \cite{Femmer20172}. 
 We did our best to follow all instructions mentioned by Femmer et al. \cite{Femmer20172} to make a free implementation of Smella, to be used for evaluation of the requirements defined in English. The original Smella tool has been evaluated on four projects for which the requirements are written in the German language. The first three cases contain requirements produced in different industrial contexts, and the fourth one is a student project.

We followed the very same process described by Femmer et al. in Section 4.2 of their article \cite{Femmer20172} to reimplement their tool. They have used hand-made dictionaries based on the ISO/IEC/IEEE 29148:2011 standard \cite{ISO/IEC/IEEE20112} to detect requirements smells.
The ISO/IEC/IEEE 29148:2011 standard defines the construct of a good requirement and provides attributes and characteristics of requirements. This standard also provides guidelines to characterize a requirement definition and offer some examples for each language criterion. Requirements language criteria are described in Section 5.2.7 of the ISO/IEC/IEEE 29148:2011 standard document \cite{ISO/IEC/IEEE20112}.
We used the examples provided by the standard and searched the WordNet dictionary for finding their synsets to manually build dictionaries similar to the ones in reference \cite{Femmer20172}. We compared the performance of their hand-made dictionary with our automatically-built dictionary on the prepared dataset.

\subsubsection{Evaluation metrics}
\noindent
To measure the performance of the proposed method in finding requirement smells, we use F1 score, Precision, and Recall given by equations (\ref{eq:F1}) to (\ref{eq:Recall}).
These are standard and widely used metrics for evaluating the performance of binary classification models, such as our neural word-embedding technique for detecting smelly or context-sensitive words. Accuracy measures the overall correctness of the model, precision measures the proportion of true positives among the predicted positives, recall measures the proportion of true positives among the actual positives, and F1 score measures the harmonic mean of precision and recall. These metrics allow us to assess the effectiveness and efficiency of our smell detection technique and compare it with the state-of-the-art tool, Smella \cite{Femmer20172}.
As mentioned, the F1 score is defined as the harmonic mean of the Precision and Recall:

\begin{equation}\label{eq:F1}
    F_1\ =\ 2\times\frac{PPV\times TPR}{PPV+TPR}
\end{equation}

\noindent where PPV is the positive predictive rate known as Precision, determining the fraction of relevant instances among all retrieved instances. TPR is the true positive rate, known as Recall or sensitivity. TPR determines the fraction of all relevant retrieved instances.
 Precision and Recall are given by the following equations:
 
\begin{equation}\label{eq:Precision}
    PPV\ =\ \frac{TP}{TP+FP\ }   
\end{equation}

\begin{equation}\label{eq:Recall}
    TPR\ =\ \frac{TP}{TP+FN\ }
\end{equation}

In the above relations, TP indicates the number of detected smelly words; FP indicates the number of words that do not have any smells but are selected incorrectly as smelly. Finally, FN is the number of smelly words that are not detected as smelly.

To compute requirements testability based on the number and type of the requirements smells, we use Equation \ref{eq:4}, which computes testability in the range (0, 1]. Errors in our automatic evaluation of testability are calculated as the difference between the testability values obtained automatically and values computed manually (ground truth values). 
To evaluate errors in the automatic computation of requirements testability, we report standard error metrics, including mean absolute error (MAE), mean squared error (MSE), root means square error (RMSE), mean squared logarithmic error (MSLnE), and median absolute error (MdAE).
Mean absolute error measures the average magnitude of the errors in the predictions, and mean square error measures the average squared magnitude of the errors in the predictions. These metrics allow us to evaluate the accuracy and robustness of our testability prediction model and demonstrate its potential for practical use.
Error metrics are computed by Equations (\ref{eq:9}) to (\ref{eq:12}). In these equations, $\ {\hat{y}}_i$ indicates the value computed automatically for the ith requirement sample; $y_i$ is the true value for ${\hat{y}}_i$, computed manually;  $y = <y_1,\ \ldots,\ y_n>$, and $n$ is the number of the samples.

\begin{equation}\label{eq:9}
    \mathrm{MAE}\left(y,\hat{y}\right)=\frac{1}{n}\sum_{i=0}^{n-1}{\left|y_i-\widehat{y_i}\right|}
\end{equation}

\begin{equation}\label{eq:10}
    MSE((y,\ \hat{y})\ =\ \frac{1}{n}\sum_{i=1}^{n-1}{{(y_i-{\hat{y}}_i)}^2}
\end{equation}

\begin{equation}\label{eq:11}
    \mathrm{MSLnE}\left(y,\hat{y}\right)=\frac{1}{n}\sum_{i=0}^{n-1}\left(\log_e{\left(1+y_i\right)}-\log_e{\left(1+{\hat{y}}_i\right)}\right)^2
\end{equation}

\begin{equation}\label{eq:12}
    \mathrm{MdAE}\left(y,\hat{y}\right)=\mathrm{median(|}y_1-\widehat{y}_1\mathrm{|, \ . . .  ,|}y_n-\widehat{y}_n\mathrm{|)}
\end{equation}

\subsection{Requirement smell prevalence}
\begin{figure}[!h]
    \centering
    \includegraphics[width=0.70\linewidth]{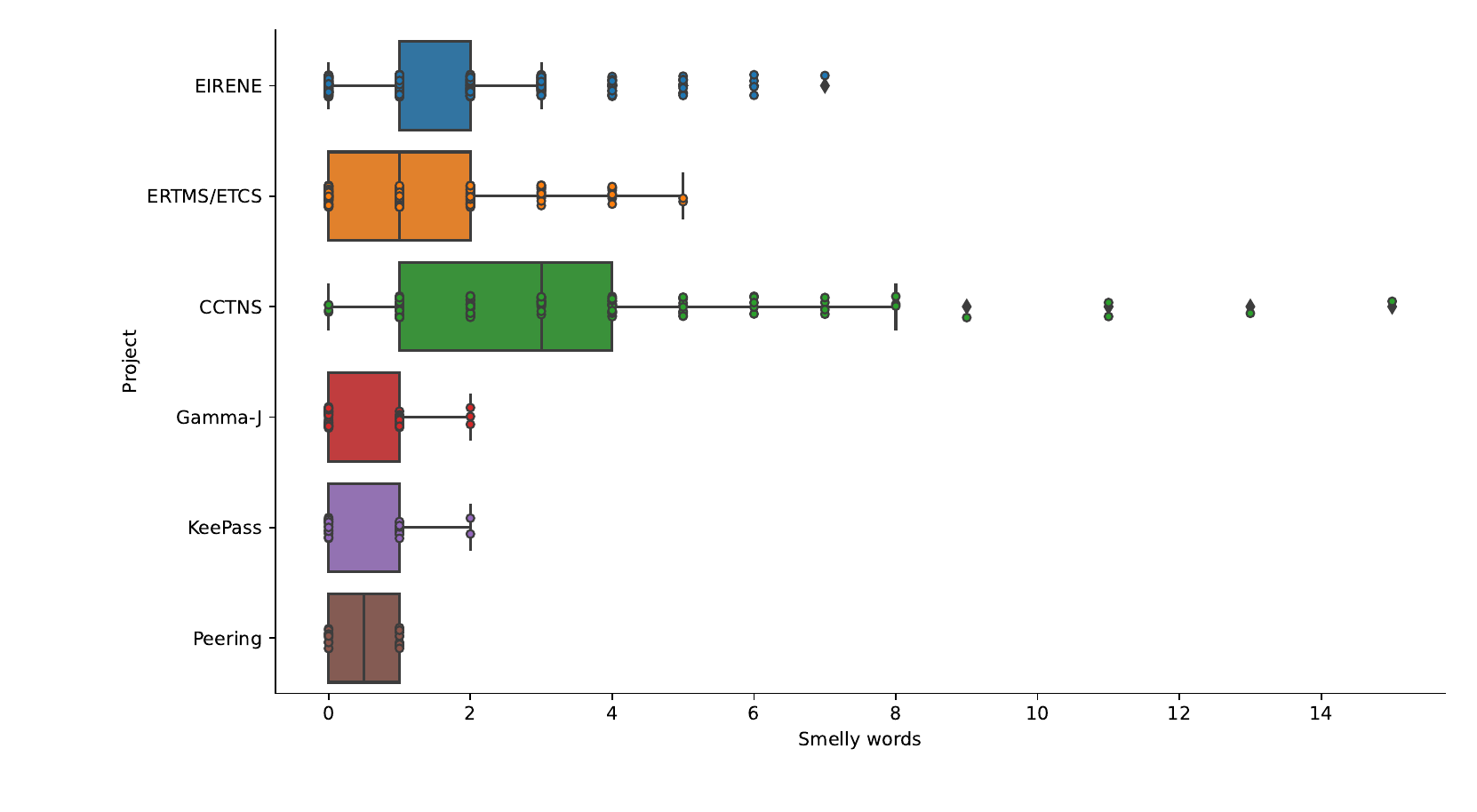}
    \caption{Frequency of the requirement smells}
    \label{fig:requirement-smells-frequency}
\end{figure}

\begin{figure}[!h]
    \centering
    \includegraphics[width=1\linewidth]{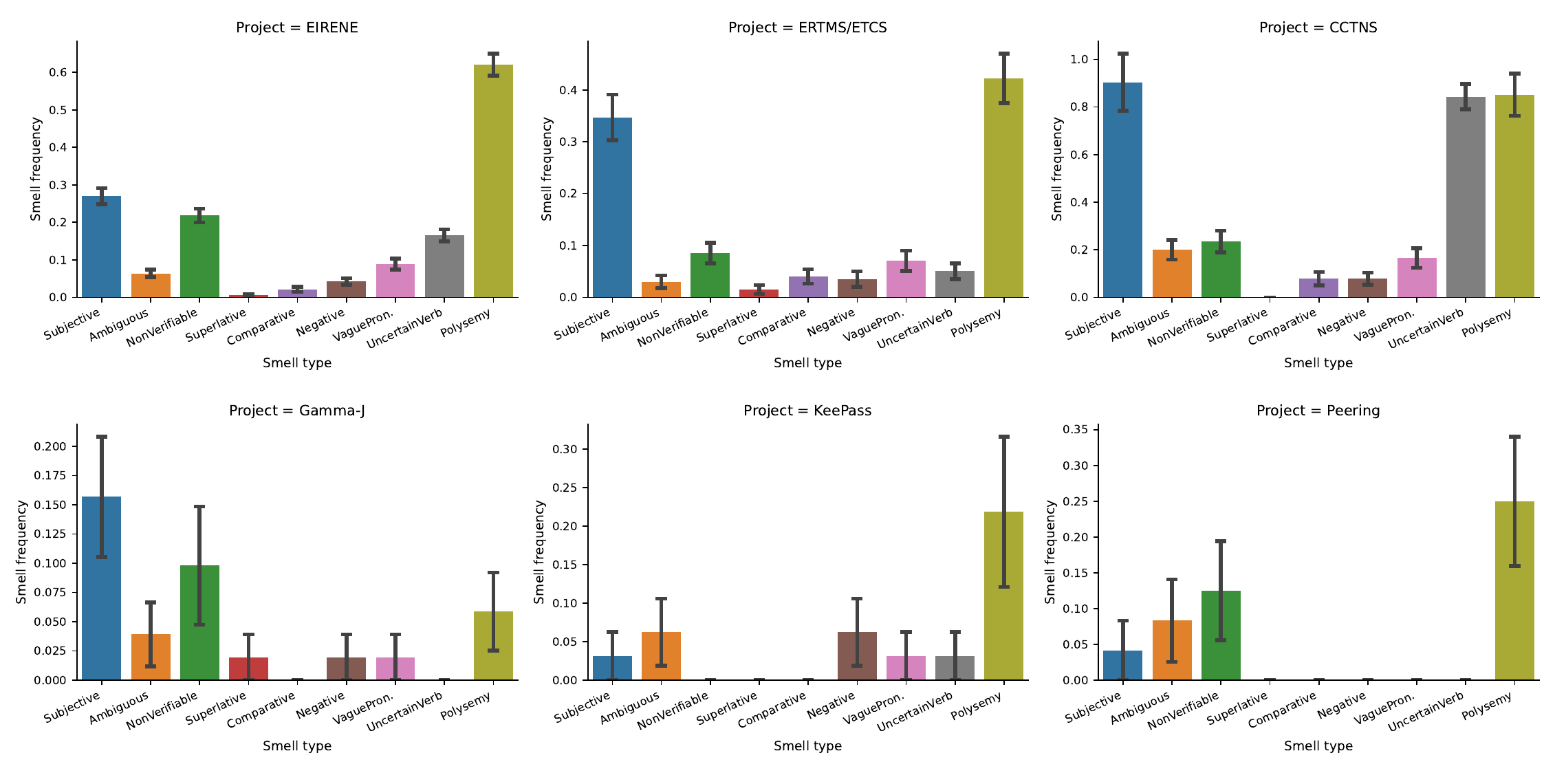}
    \caption{Variation of smells within the requirements of each project}
    \label{fig:requirement-smells-variation}
\end{figure}

\noindent
To answer RQ\textsubscript{1}, we measure the frequency of smells in our labeled dataset. Figure \ref{fig:requirement-smells-frequency} shows the box plot of existing smells in each project's requirements. Datapoints denote the number of smells in each requirement. It is observed that CCTNS has the most smells, while the number of the requirement of this project is significantly lower than the EIRENE project. Therefore, it is observed that the frequency of smells does not depend on the number of requirements in the project. 
Figure \ref{fig:requirement-smells-variation} shows the frequency of the requirements smells types. For each project, the mean of smells' frequency along with the confidence interval of 95\% for mean value, $\mu$, around bootstrapped values \cite{Diciccio1996}, is shown by error bars ($\updownarrow$) in Figure \ref{fig:requirement-smells-variation}.

One can observe that Polysemy and Subjective language are the two most prevalent smells in requirements. Non-verifiable terms are also frequent; nevertheless, other smells occur occasionally. On average, the fraction of smelly words to the total requirement's words in our dataset is about 4.93\%, which indicates that the frequency of requirements smells are \emph{similar} to the code smells \cite{Pecorelli2020}. However, the percentage of requirements with at least one smell is 72.39\%, denoting that most of the textual requirements suffer from smells and need to modify or refactor before use. Therefore, automated mechanisms to find requirement smells are necessary and beneficial to the SDLC. 

\begin{figure}
    \centering
    \includegraphics[width=0.60\linewidth]{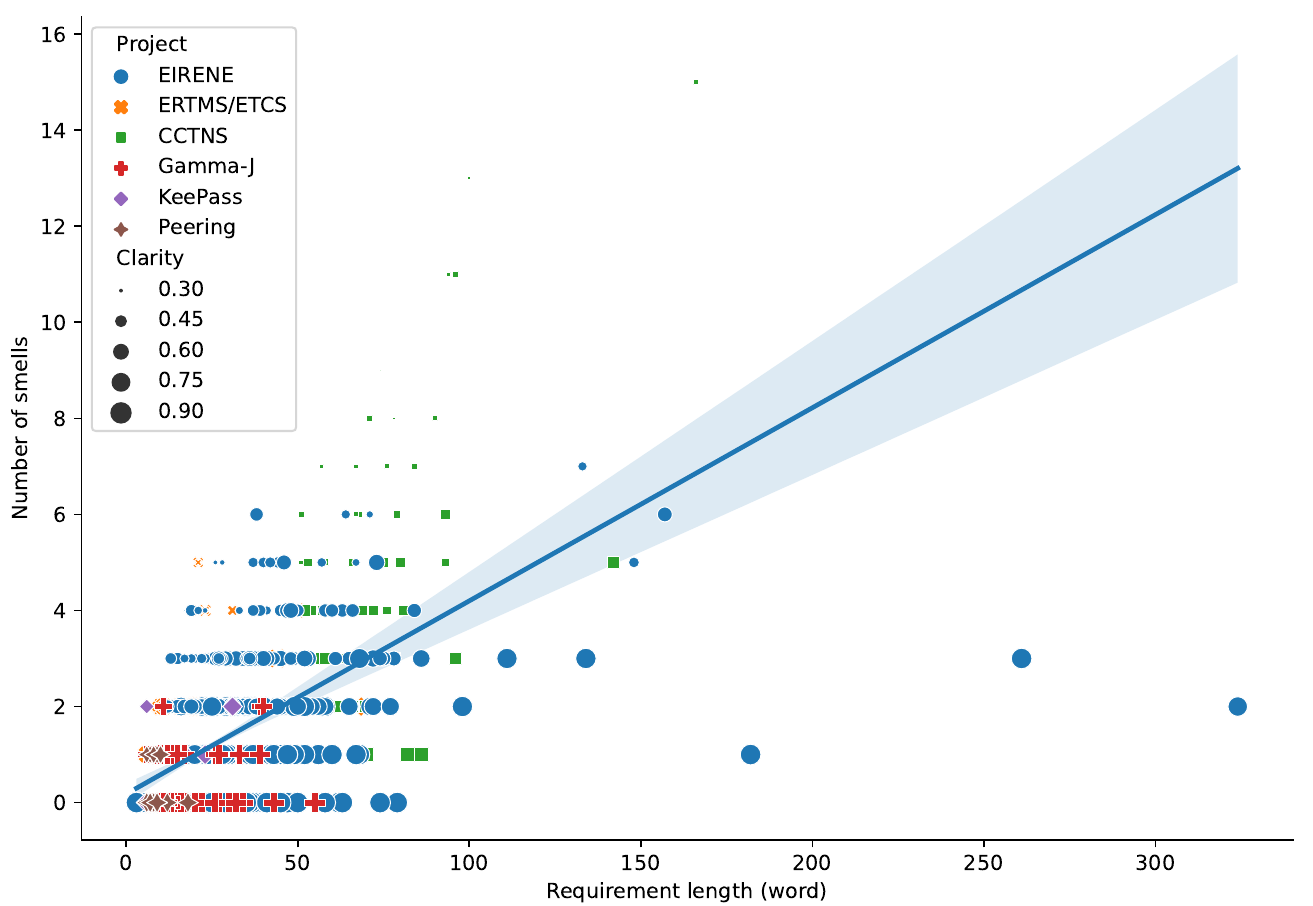}
    \caption{Number of smells against the requirement size along with a linear regression line}
    \label{fig:number-of-smells-against-requirement-size}
\end{figure}

Figure \ref{fig:number-of-smells-against-requirement-size} shows the number of smells in each requirement against the requirement's length, \emph{i.e.}, the number of words. The continuous line illustrates the regression line between the two variables, and the bulleted numbers denote the requirements clarity degree.
It can be observed that as the number of words in a requirement increases, the number of smells increases. 
The Spearman rank-order correlation coefficient between the number of smells and the length of the requirements for our dataset is $0.5253$ with a p-value of $5.9998\ \times{10}^{-71}$, which means that they are positively correlated. 
As we expected, the requirements with smaller lengths are cleaner and more testable than requirements with many words.

A relatively similar result has been proposed by Femmer et al. \cite{Femmer20172}, where a Spearman correlation of $0.9$ between the number of smells and the length of the requirements has been reported. Their result shows that the number of detected smells strongly correlates with the size of the requirement artifact, while for our study, the correlation is not very strong. 
One primary reason is the language used to express the requirements. It seems that English is more suitable to express requirements than German. Another reason is that Femmer et al. \cite{Femmer20172} use their findings to compute the correlation coefficient, which is prone to be a wrong result due to the low precision of automatic smell detection for most types of smells. Hence, we believe that our results are more realistic than the results reported by Femmer et al. \cite{Femmer20172}.

\begin{tcolorbox}
    \textbf{Answer to RQ\textsubscript{1}}:
    \emph{
    Only 4.93\% of words are recognized as smelly words in the software requirements, while 72.39\% of requirements contain at least one kind of smell. Polysemy, Subjective language, and Non-verifiable terms are the most prevalent requirement smells. The number of smells increases as the requirement length grows, with a correlation coefficient of $0.5253$.
}
\end{tcolorbox}

\subsection{Smelly words dictionary analysis}\label{sec:smelly-words-dictionary-analysis}
\noindent
The descriptive analysis of the requirements dataset in the previous section shows that the most prevalent requirement smells, \emph{i.e.}, Polysemy, Subjective language, and Non-verifiable terms are smells whose current detection mechanism is based on a dictionary. Therefore, it is essential to build an efficient and comprehensive dictionary to harness these smells. 
The first 30 words and the last five words of the automated dictionary with corresponding  similarity values and smells are shown in Table \ref{table:smelly-words-dictionary}. The dictionary contains the top 1,000 frequent words in computer science sorted in descending order by their similarity in 10 domains. The "mean" similarity column shows the average cosine similarity between each word in computer science and other domains.

Words with different meanings in different domains were less similar. For example, the word "call" has the least similarity since it has various meanings, such as "telephone call", "voice", and "entitle". At the same time, in computer science, the word "call" usually means "invoking a function", "method", or "subroutine" of the program, which is different from the other domains. 
In such cases, according to the context, the requirement engineer can replace the smelly word with non-smelly and clear ones. For instance, the word "call" may be replaced with "phone call" or "function call" in a requirement definition.
In the last rows, the words that have a single and exact meaning are observed. These words are mostly specific to the computer science domain. For example, "object-oriented", "c++", and "boolean" are dedicated words in computer science, even when used in other domains. Therefore, they showed high similarity for all the predefined domains.

After ranking each word based on its similarity measure, a thorough investigation was made to label each word with its exact type of smell. 
Sorting the list of the ranked words, on the similarity measure, we noticed that the words with a mean cosine similarity of greater than or equal to 0.5943 do not expose any smell. Specifically, the word “association” with a mean cosine similarity of 0.5943 is the last smelly word in our sorted list, shown in Table \ref{table:smelly-words-dictionary}. 

We observed that words with a similarity above 0.5943 are specific to the computer science domain or represent only one concept. Therefore, these words (about 250 words of 1,000 preliminary words) do not indicate any smell.
It is worth noting that there are words with a low similarity value that have no smells. For example, the word "add" is a verb identified as a clean word by our experts. Nevertheless, the proposed mechanism for selecting candidate words and computing words' similarity with the Word2Vec algorithm sounds efficient and applicable to any language. 

  \begin{table}[!h]
     \centering
     \caption{Smelly words dictionary with labels and similarity values}
     \label{table:smelly-words-dictionary}
     \resizebox{1\linewidth}{!}{%
         \begin{tabular}{lrrrrrrrrrrrl} 
             \hline
             \begin{tabular}[c]{@{}l@{}}Frequent \\CS word\end{tabular} & \multicolumn{1}{l}{SS} & \multicolumn{1}{l}{Lw} & \multicolumn{1}{l}{Ec} & \multicolumn{1}{l}{Ci} & \multicolumn{1}{l}{At} & \multicolumn{1}{l}{Li} & \multicolumn{1}{l}{EE} & \multicolumn{1}{l}{ME} & \multicolumn{1}{l}{Sp} & \multicolumn{1}{l}{Me} & \multicolumn{1}{l}{Mean}       & Smell                 \\ 
             \hline
             call                                          & 0.2556                 & 0.0993                 & -0.0076                & 0.1854                 & 0.0513                 & 0.1037                 & 0.1099                 & 0.2191                 & 0.0216                 & 0.0393                 & \textbf{0.1078}                & Polysemy              \\
             part                                          & 0.2612                 & 0.1506                 & 0.0837                 & 0.1368                 & 0.0708                 & 0.2082                 & 0.1674                 & 0.1280                 & 0.1409                 & 0.1551                 & \textbf{0.1503}                & Subjective-language   \\
             present                                       & 0.2415                 & 0.1947                 & 0.2045                 & 0.1286                 & 0.1191                 & 0.1147                 & 0.1536                 & 0.1917                 & 0.0799                 & 0.2485                 & \textbf{0.1677}                & Polysemy              \\
             another                                       & 0.2770                 & 0.1948                 & 0.1912                 & 0.0790                 & 0.1932                 & 0.2597                 & 0.1489                 & 0.1612                 & 0.0397                 & 0.1397                 & \textbf{0.1684}                & Non-verifiable terms  \\
             include                                       & 0.1852                 & 0.1677                 & 0.0598                 & 0.1489                 & 0.1586                 & 0.2206                 & 0.2656                 & 0.2754                 & 0.1140                 & 0.1091                 & \textbf{0.1705}                & Non-verifiable terms  \\
             along                                         & 0.1732                 & 0.1975                 & 0.1285                 & 0.2035                 & 0.1185                 & 0.2308                 & 0.2435                 & 0.2244                 & 0.0241                 & 0.1743                 & \textbf{0.1718}                & Ambiguous Adv./Adj.   \\
             take                                          & 0.3570                 & 0.1006                 & 0.1204                 & 0.1772                 & 0.2616                 & 0.2125                 & 0.1474                 & 0.2900                 & 0.0813                 & 0.1972                 & \textbf{0.1945}                & Polysemy              \\
             already                                       & 0.2950                 & 0.1905                 & 0.1267                 & 0.1750                 & 0.2094                 & 0.1838                 & 0.2782                 & 0.2716                 & 0.1543                 & 0.1264                 & \textbf{0.2011}                & Ambiguous Adv./Adj.   \\
             entire                                        & 0.4104                 & 0.2050                 & 0.1224                 & 0.2752                 & 0.2008                 & 0.2056                 & 0.2246                 & 0.2423                 & 0.0481                 & 0.0913                 & \textbf{0.2026}                & Subjective-language   \\
             full                                          & 0.2906                 & 0.0909                 & 0.0786                 & 0.2157                 & 0.1810                 & 0.2097                 & 0.3246                 & 0.3175                 & 0.1624                 & 0.1849                 & \textbf{0.2056}                & Non-verifiable terms  \\
             add                                           & 0.2838                 & 0.1889                 & 0.0781                 & 0.0831                 & 0.2910                 & 0.1506                 & 0.3537                 & 0.2867                 & 0.1702                 & 0.1844                 & \textbf{0.2070}                & -                     \\
             originally                                    & 0.2812                 & 0.1603                 & 0.0463                 & 0.2590                 & 0.1833                 & 0.1929                 & 0.3361                 & 0.3191                 & 0.1930                 & 0.2083                 & \textbf{0.2180}                & Ambiguous Adv./Adj.   \\
             previous                                      & 0.2815                 & 0.2184                 & 0.2743                 & 0.2052                 & 0.2230                 & 0.2479                 & 0.2319                 & 0.2607                 & 0.1370                 & 0.2157                 & \textbf{0.2296}                & Ambiguous Adv./Adj.   \\
             main                                          & 0.3090                 & 0.1158                 & 0.1475                 & 0.2320                 & 0.2468                 & 0.2231                 & 0.2672                 & 0.3294                 & 0.1871                 & 0.2586                 & \textbf{0.2317}                & Polysemy              \\
             similar                                       & 0.3339                 & 0.1896                 & 0.1181                 & 0.2023                 & 0.1978                 & 0.2594                 & 0.3135                 & 0.3294                 & 0.1580                 & 0.2424                 & \textbf{0.2344}                & Subjective-language   \\
             also                                          & 0.3155                 & 0.1537                 & 0.2034                 & 0.2222                 & 0.2968                 & 0.2697                 & 0.2685                 & 0.2713                 & 0.1552                 & 0.2152                 & \textbf{0.2372}                & -                     \\
             made                                          & 0.3313                 & 0.2491                 & 0.1927                 & 0.1541                 & 0.2650                 & 0.3075                 & 0.1972                 & 0.2983                 & 0.1816                 & 0.2025                 & \textbf{0.2379}                & Polysemy              \\
             third                                         & 0.2974                 & 0.2196                 & 0.1359                 & 0.2283                 & 0.2091                 & 0.2262                 & 0.2541                 & 0.2974                 & 0.2024                 & 0.3638                 & \textbf{0.2434}                & -                     \\
             whose                                         & 0.3652                 & 0.2408                 & 0.1707                 & 0.2232                 & 0.2675                 & 0.2951                 & 0.2259                 & 0.3819                 & 0.2005                 & 0.1109                 & \textbf{0.2482}                & Ambiguous Adv./Adj.   \\
             start                                         & 0.2834                 & 0.2368                 & 0.1147                 & 0.2395                 & 0.2582                 & 0.3080                 & 0.3353                 & 0.3685                 & 0.1701                 & 0.1957                 & \textbf{0.2510}                & Polysemy              \\
             come                                          & 0.4019                 & 0.2010                 & 0.1851                 & 0.1940                 & 0.2852                 & 0.2377                 & 0.2728                 & 0.2858                 & 0.2359                 & 0.2168                 & \textbf{0.2516}                & -                     \\
             section                                       & 0.3110                 & 0.3493                 & 0.1927                 & 0.2092                 & 0.0943                 & 0.2678                 & 0.2992                 & 0.3845                 & 0.1160                 & 0.2971                 & \textbf{0.2521}                & Polysemy              \\
             base                                          & 0.4224                 & 0.2937                 & 0.2077                 & 0.2849                 & 0.1788                 & 0.2559                 & 0.3755                 & 0.3005                 & 0.0944                 & 0.1362                 & \textbf{0.2550}                & Polysemy              \\
             together                                      & 0.3719                 & 0.2903                 & 0.2291                 & 0.1992                 & 0.2872                 & 0.1722                 & 0.2503                 & 0.3338                 & 0.2420                 & 0.1949                 & \textbf{0.2571}                & Subjective-language   \\
             long                                          & 0.2112                 & 0.1932                 & 0.1269                 & 0.2155                 & 0.3180                 & 0.4016                 & 0.2853                 & 0.3117                 & 0.3016                 & 0.2152                 & \textbf{0.2580}                & Non-verifiable terms  \\
             initial                                       & 0.3660                 & 0.3135                 & 0.2138                 & 0.1846                 & 0.1986                 & 0.2404                 & 0.3449                 & 0.3607                 & 0.1596                 & 0.2050                 & \textbf{0.2587}                & Polysemy              \\
             following                                     & 0.2980                 & 0.2050                 & 0.2668                 & 0.2162                 & 0.2681                 & 0.3331                 & 0.2942                 & 0.3092                 & 0.2107                 & 0.2070                 & \textbf{0.2608}                & Non-verifiable terms  \\
             known                                         & 0.3065                 & 0.2525                 & 0.2288                 & 0.2375                 & 0.2039                 & 0.3220                 & 0.2796                 & 0.3481                 & 0.1955                 & 0.2803                 & \textbf{0.2655}                & Subjective-language   \\
             left                                          & 0.3595                 & 0.1969                 & 0.1169                 & 0.1753                 & 0.2929                 & 0.2991                 & 0.3343                 & 0.3560                 & 0.2588                 & 0.2750                 & \textbf{0.2665}                & Polysemy              \\
             …                                             & \multicolumn{1}{l}{…}  & \multicolumn{1}{l}{…}  & \multicolumn{1}{l}{…}  & \multicolumn{1}{l}{…}  & \multicolumn{1}{l}{…}  & \multicolumn{1}{l}{…}  & \multicolumn{1}{l}{…}  & \multicolumn{1}{l}{…}  & \multicolumn{1}{l}{…}  & \multicolumn{1}{l}{…}  & \multicolumn{1}{l}{\textbf{…}} & …                     \\
             \rowcolor[rgb]{0.937,0.937,0.937} association & 0.6754                 & 0.5938                 & 0.5140                 & 0.5576                 & 0.4982                 & 0.5582                 & 0.6628                 & 0.7138                 & 0.5486                 & 0.6205                 & \textbf{0.5943}                & Polysemy              \\
             …                                             & \multicolumn{1}{l}{…}  & \multicolumn{1}{l}{…}  & \multicolumn{1}{l}{…}  & \multicolumn{1}{l}{…}  & \multicolumn{1}{l}{…}  & \multicolumn{1}{l}{…}  & \multicolumn{1}{l}{…}  & \multicolumn{1}{l}{…}  & \multicolumn{1}{l}{…}  & \multicolumn{1}{l}{…}  & \multicolumn{1}{l}{\textbf{…}} & …                     \\
             mathematician                                 & 0.8688                 & 0.7527                 & 0.7337                 & 0.7666                 & 0.7550                 & 0.7832                 & 0.8694                 & 0.8943                 & 0.8276                 & 0.8265                 & \textbf{0.8078}                & -                     \\
             campus                                        & 0.8890                 & 0.8208                 & 0.8110                 & 0.5987                 & 0.8450                 & 0.8634                 & 0.8481                 & 0.8758                 & 0.7585                 & 0.8185                 & \textbf{0.8129}                & -                     \\
             psychology                                    & 0.9075                 & 0.8486                 & 0.8005                 & 0.7675                 & 0.7624                 & 0.7466                 & 0.8784                 & 0.8613                 & 0.7558                 & 0.8884                 & \textbf{0.8217}                & -                     \\
             transistor                                    & 0.9181                 & 0.8378                 & 0.8487                 & 0.7249                 & 0.8515                 & 0.8016                 & 0.9454                 & 0.9250                 & 0.7137                 & 0.8169                 & \textbf{0.8384}                & -                     \\
             bachelor                                      & 0.9451                 & 0.8951                 & 0.7462                 & 0.8573                 & 0.7746                 & 0.7050                 & 0.9391                 & 0.9319                 & 0.7910                 & 0.8417                 & \textbf{0.8427}                & -                     \\
             nobel                                         & 0.9466                 & 0.8532                 & 0.8438                 & 0.7646                 & 0.7467                 & 0.7315                 & 0.9549                 & 0.9516                 & 0.7962                 & 0.8593                 & \textbf{0.8448}                & -                     \\
             boolean                                       & 0.7968                 & 0.7875                 & 1.0000                 & 1.0000                 & 0.7515                 & 0.5891                 & 0.8090                 & 0.7170                 & 1.0000                 & 1.0000                 & \textbf{0.8451}                & -                     \\
             undergraduate                                 & 0.9233                 & 0.8025                 & 0.8521                 & 0.7699                 & 0.7883                 & 0.8693                 & 0.8950                 & 0.8924                 & 0.8442                 & 0.8787                 & \textbf{0.8516}                & -                     \\
             c++                                           & 0.8431                 & 0.7483                 & 1.0000                 & 1.0000                 & 0.6741                 & 0.7058                 & 0.8694                 & 0.7278                 & 1.0000                 & 1.0000                 & \textbf{0.8568}                & -                     \\
             object-oriented                               & 0.8209                 & 0.6667                 & 1.0000                 & 1.0000                 & 0.7013                 & 1.0000                 & 0.8233                 & 0.7220                 & 1.0000                 & 1.0000                 & \textbf{0.8734}                & -                     \\
             \textbf{Mean}                                 & \textbf{0.6288}        & \textbf{0.4997}        & \textbf{0.4404}        & \textbf{0.4404}        & \textbf{0.4974}        & \textbf{0.4920}        & \textbf{0.6190}        & \textbf{0.6011}        & \textbf{0.4531}        & \textbf{0.4970}        & \textbf{0.5169}                & \textbf{-} \\ \hline          
         \end{tabular}
     }
 \end{table}

\begin{figure}
    \centering
    \includegraphics[width=0.95\linewidth]{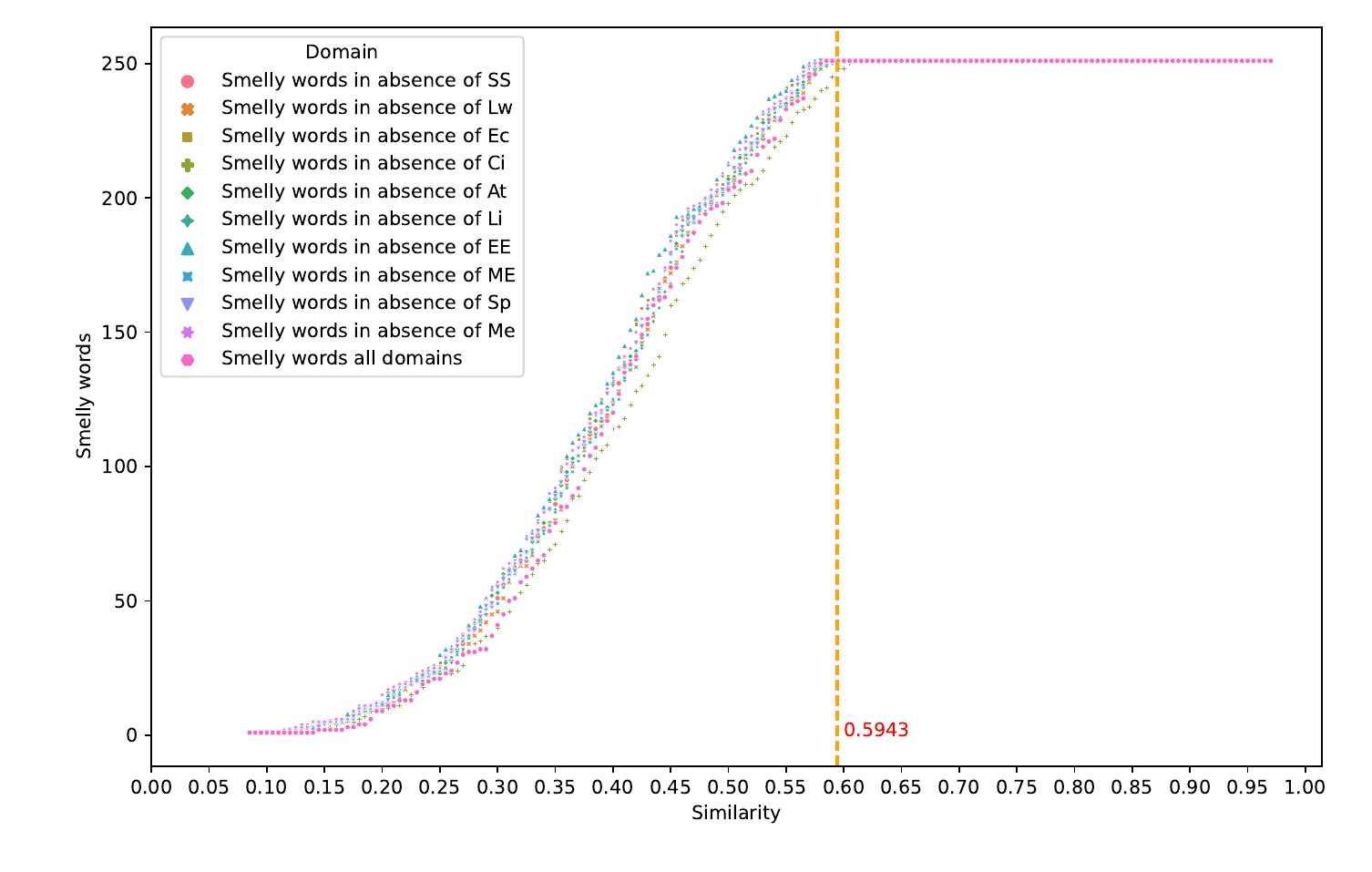}
    \caption{Number of smelly words for different similarity values}
    \label{fig:auto-dict-scatter-plot}
\end{figure}

We performed a sensitivity analysis \cite{Frey2002} to show that the threshold identification is robust and domain-independent. We repeated the calculation of mean similarity for the top 1,000 frequent words in computer science ten times. For each experiment, one application domain was removed, and the mean similarity was calculated to measure the impact of the removed domain on the obtained threshold. 
Figure \ref{fig:auto-dict-scatter-plot} shows a scatter plot of the cumulative number of smelly words at each similarity point. The total number of smelly words reaches a constant value as the similarity exceeds 0.5943 in each experiment. Indeed, eliminating one application domain does not change the number of smelly words in our experiments. 

Table \ref{table:dictionary-sensitivity-analysis}, for each experiment, shows the changes in the mean similarity of the last smelly word in the list of the words, sorted on the order of their similarity with the 1,000 most frequent computer science words. We observe small changes (less than 0.0057 or 5.7\% on average) in the similarity threshold compared to the experiment that takes all domains into account. Therefore, it is concluded that the obtained threshold is not sensitive to any specific application domain. It is worth mentioning that frequent words in the computer whose similarity measure is greater than the observed threshold are not smelly, while those with a similarity measure less than the threshold are mostly smelly.

\begin{table}[!h]
    \centering
    \caption{Variations in the similarity of the last smelly words in the automatically created dictionary}
    \label{table:dictionary-sensitivity-analysis}
    \resizebox{0.80\linewidth}{!}{%
        \begin{tabular}{lll} 
            \hline
            Domains                                       & The similarity of the last smelly word & Absolute change in similarity value  \\ 
            \hline
            \rowcolor[rgb]{0.886,0.937,0.851} All domains & 0.5943                                 & 0                                    \\
            All domains except SS                         & 0.5943                                 & 0                               \\
            All domains except Lw                         & 0.5943                                 & 0                               \\
            All domains except Ec                         & 0.6032                                 & 0.0089                               \\
            All domains except Ci                         & 0.5984                                 & 0.0041                               \\
            All domains except At                         & 0.6050                                 & 0.0107                              \\
            All domains except Li                         & 0.5983                               & 0.0040                               \\
            All domains except EE                         & 0.5867                                 & 0.0076                               \\
            All domains except ME                         & 0.5810                                 & 0.0133                               \\
            All domains except Sp                         & 0.5994                                & 0.0051                               \\
            All domains except Me                         & 0.5913                                 & 0.0030                               \\
            Average                                       & --                                  & 0.0057                              \\
            \hline
        \end{tabular}
    }
\end{table}

\begin{tcolorbox}
    \textbf{Answer to RQ\textsubscript{2}}:
    \emph{
        We observed that words, such as "call", "part", and "present", with an average cosine similarity of less than $0.5943$ in all domains, are most likely to have different meanings in different domains of use. On the other hand, words such as "object-oriented", "c++", and "cpu" with an average cosine similarity close to one, are most likely clear and unambiguous, without any smells.
    }
\end{tcolorbox}

\subsection{Requirement smell detection analysis}
\noindent
To evaluate the proposed method's effectiveness in detecting requirement smells, we ran Algorithm \ref{alg:2} on our dataset and compared the results with manually labeled data as the ground truth. 
As discussed earlier, for uncertain verbs, we use a small hand-made dictionary consisting of common English modals: "may", "might", " can", " could", and" should". 
 We made an exception for the word “shall” as it is often used to express functional requirements \cite{Wiegers2013}.
For other smells, identified by a dictionary mechanism, we used the dictionary of smelly words that was prepared automatically. We also implemented, Smella \cite{Femmer20172} to compare its performance with our smell detection tool. 

Table \ref{table:smell-detection-performance} shows the Precision, Recall, and F1 for detecting different types of smells for all the requirements in our dataset.
It is observed that for the Polysemy smell, as the most prevalent requirement smell, both the Precision and Recall are quite remarkable and demonstrate the applicability of our smell detection method in practice. We demonstrate the applicability of our approach on several examples of real-world software requirements in Section  \ref{sec:qualitative-analysis}.

The gap between Precision and Recall is very high for most smell types. Indeed, the Recall is higher than the Precision, which means that our algorithm is good enough to identify all smelly words, but it also selects words that do not contain any smell and hence leads to a high \emph{false-positive rate}. 
The primary reason is that the words are only detected based on their grammatical role or their existence in a predefined dictionary, which may be misleading for a particular context. It can be realized that reducing the false positive rate in requirement smell detection tools needs more advanced techniques, considering the overall context of the requirement statement. 

\begin{figure}
    \centering
    \includegraphics[width=1\linewidth]{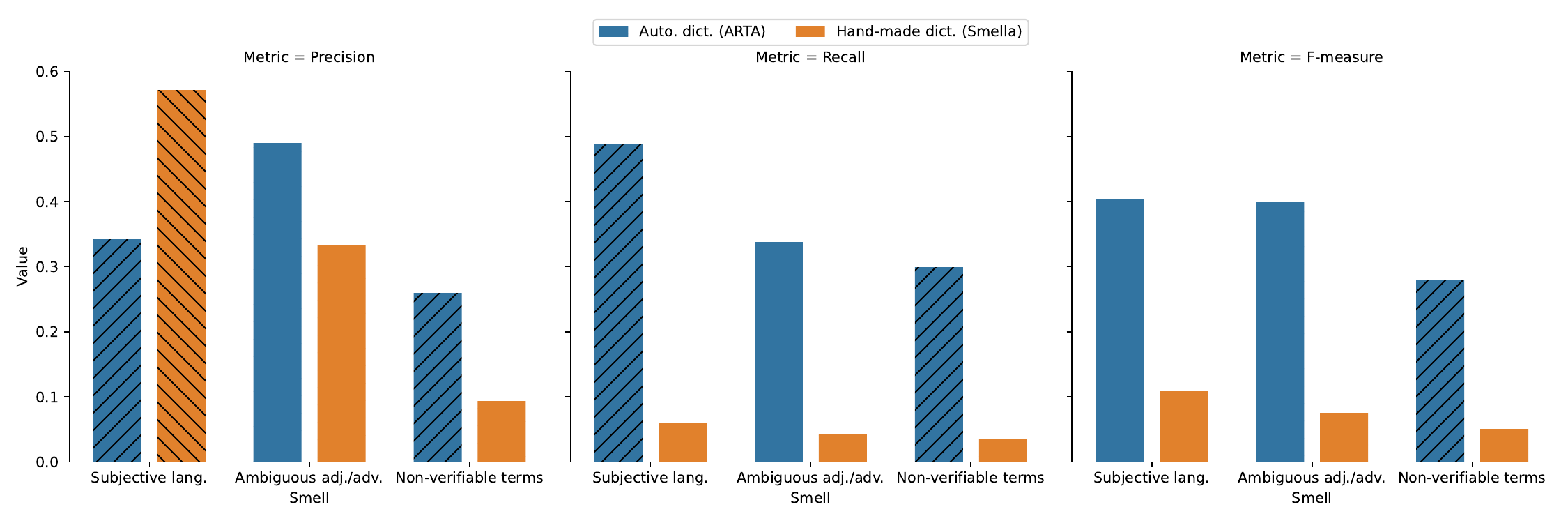}
    \caption{Performance of ARTA and Smella \cite{Femmer20172} in detecting dictionary-based requirement smells}
    \label{fig:arta-vs-smella}
\end{figure}

The smells with lower F1 scores are, apparently, more difficult to detect. For example, the F1 score of vague pronouns is very low since determining the references for all the pronouns in a sentence is a complicated and error-prone NLP task. However, as shown in Figure \ref{fig:arta-vs-smella}, our approach outperforms the state-of-the-art requirements smell detection tool, \emph{i.e.}, Smella \cite{Femmer20172}. 
The figure shows the Precision, Recall, and F1 score of ARTA and Smella for three types of requirement smells detected with the dictionary. Our approaches have a better performance than Smella in all three smells. The Smella tool could not recognize Polysemy smells. Therefore, we have not illustrated the performance of this smell in Figure \ref{fig:arta-vs-smella}. 
ARTA improves the Precision, Recall, and F1 score of dictionary-based smell detection, respectively, by an average of 0.0311, 0.3299, 0.2824.

\begin{table}
    \centering
    \caption{Smell detection performance}
    \label{table:smell-detection-performance}
    \resizebox{0.45\linewidth}{!}{%
        \begin{tabular}{llll}
            \hline
            Requirement smell    & Precision & Recall & F1 \\ \hline
            Subjective language  & 0.3421    & 0.4885 & 0.4024    \\
            Ambiguous adv./ adj. & 0.4898    & 0.3380 & 0.4000    \\
            Non-verifiable term  & 0.2598    & 0.2994 & 0.2782    \\
            Superlative          & 0.3500    & 0.8750 & 0.5000    \\
            Comparative          & 0.4746    & 0.8750 & 0.6154    \\
            Negative             & 0.2897    & 0.9333 & 0.4421    \\
            Vague pron.          & 0.1363    & 0.6966 & 0.2279    \\
            Uncertain-verb       & 0.7766    & 0.9815 & 0.8671    \\
            Polysemy             & 0.6688    & 0.8322 & 0.7416    \\
            \textbf{Average}             & \textbf{0.4209}    & \textbf{0.7022} & \textbf{0.4972}    \\ \hline
        \end{tabular}%
    }
\end{table}

\begin{tcolorbox}
\textbf{Answer to RQ\textsubscript{3}}:
    \emph{
        On average automated dictionary improves the F-measure of requirement smell detection by 28.24\% on our dataset. Polysemy and Uncertain verbs can be detected with high Precision and Recall. On the other hand, Vague pronouns and Non-verifiable terms are more difficult to detect even when using an automated dictionary mechanism.
    }
\end{tcolorbox}

\subsection{Requirements testability analysis}\label{sec:requirement-testability-analysis}
\noindent
In our last experiment, we compute the testability value of each requirement in our dataset based on Equation \ref{eq:4}. Three computations are performed for each requirement: the first computation is based on the ground-truth dataset, the second computation is based on our approach obtained by ARTA web application, and the third computation is based on smells finding by Smella \cite{Femmer20172}. 
We reckoned the smells detected by each approach separately and calculated requirements clarity and testability values. Then, error metrics are computed by comparing the estimated value with corresponding  ground-truth values.

\subsubsection{Computing and comparing alpha values}
\noindent
Algorithm \ref{alg:2} measures requirements testability based on our testability model. 
It needs parameter alpha, the base cost of analyzing an extra sentence in a given requirement definition, discussed in Section \ref{sec:guidelines-on-determining-the-alpha-factor}.
Table \ref{table:boundary-values-of-alpha-for-all-projects} shows the computed values of alpha for all projects in our dataset. For each project, we have selected the most appropriate values for the four different aspects of alpha, shown in Figure \ref{fig:extra-sentence-cost}. The lower bound and upper bound of alpha have been determined using softened and hardened policies. 
From Table \ref{table:boundary-values-of-alpha-for-all-projects}, we can observe that the test effort, addressed by alpha, for a safety-critical project such as a train system controller, ERTMS/ETCS, is more than a password management system, KeePass.

\begin{figure}
    \centering
    \includegraphics[width=0.90\linewidth]{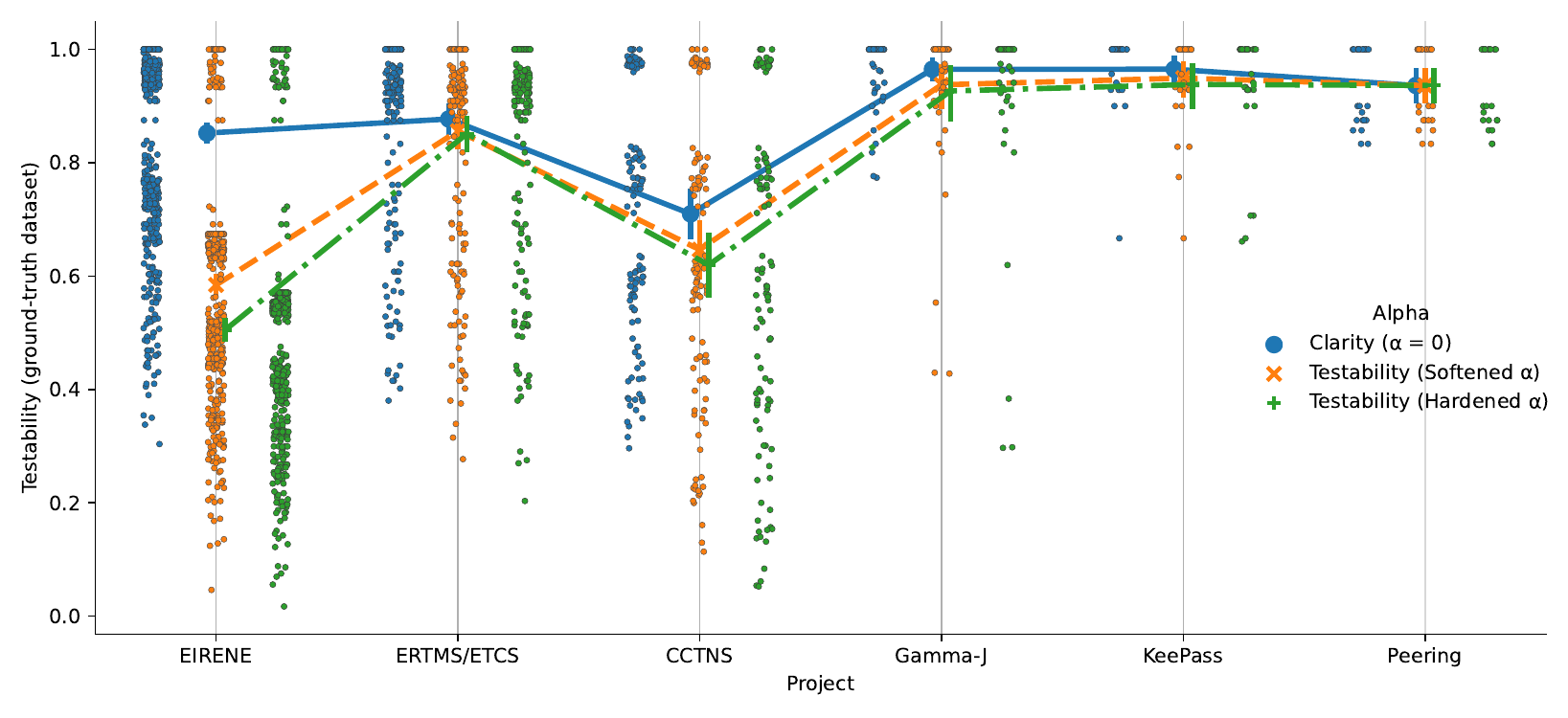}
    \caption{Requirements testability with different values of $\alpha$ for each project.}
    \label{fig:requirement-testability-comparing-alphas}
\end{figure}

Figure \ref{fig:requirement-testability-comparing-alphas} shows the impact of the parameter $\alpha$ on the requirements testability of different projects in our dataset.
The points show the value of the testability of individual requirements. The average testability of the requirements for both the softened and hardened policies of alpha is shown with different markers. The requirements testability in the absence of alpha ($\alpha=0$) is the same requirement clarity. 
Therefore, Figure \ref{fig:requirement-testability-comparing-alphas} also denotes the upper bound and lower bound for requirements testability values of each project in our dataset.
It is observed that the testability of the requirements in the EIRENE and CCTNS projects is seriously affected by parameter $\alpha$, while for other projects, testability differences are negligible. It means that most of the requirements in EIRENE and CCTNS projects, which are critical systems, have been expressed in more than one sentence.

\begin{table}
    \centering
    \caption{Boundary values of alpha for projects used in our study}
    \label{table:boundary-values-of-alpha-for-all-projects}
    \resizebox{0.85\linewidth}{!}{%
        \begin{tabular}{lllllll} 
            \hline
            \multirow{2}{*}{Project} & \multirow{2}{*}{Domain} & \multirow{2}{*}{Criticality} & \multirow{2}{*}{Type} & \multirow{2}{*}{Template} & \multicolumn{2}{c}{Alpha}                                                                  \\ 
            \cline{6-7}
            &                         &                             &                       &                           & Softened                                 & Hardened                                        \\ 
            \hline
            EIRENE                   & EE                      & Safety-critical             & Functional            & Multiple sentences        &  0.4836    &  0.7535  \\
            ERTMS/
            ETCS            & EE + ME                 & Safety-critical             & Functional            & Single sentence           &  0.6093 &  0.8792  \\
            CCTNS                    & LW                      & Business-critical           & Functional            & Multiple sentences        &  0.3102    & 0.5801  \\
            Gamma-J                  & EC + CS                 & Business-critical           & Functional            & Single sentence           &  0.3445 & 0.6144  \\
            KeePass                  & CS                      & Non-critical                & Functional            & Single sentence           &  0.2075         &  0.4150            \\
            Peering                  & CS                      & Business-critical           & Functional            & Single sentence           & 0.2700      &  0.5399       \\
            \hline
        \end{tabular}
    }
\end{table}

\subsubsection{Quantitative analysis}
\begin{figure}[!h]
    \centering
    \includegraphics[width=0.95\linewidth]{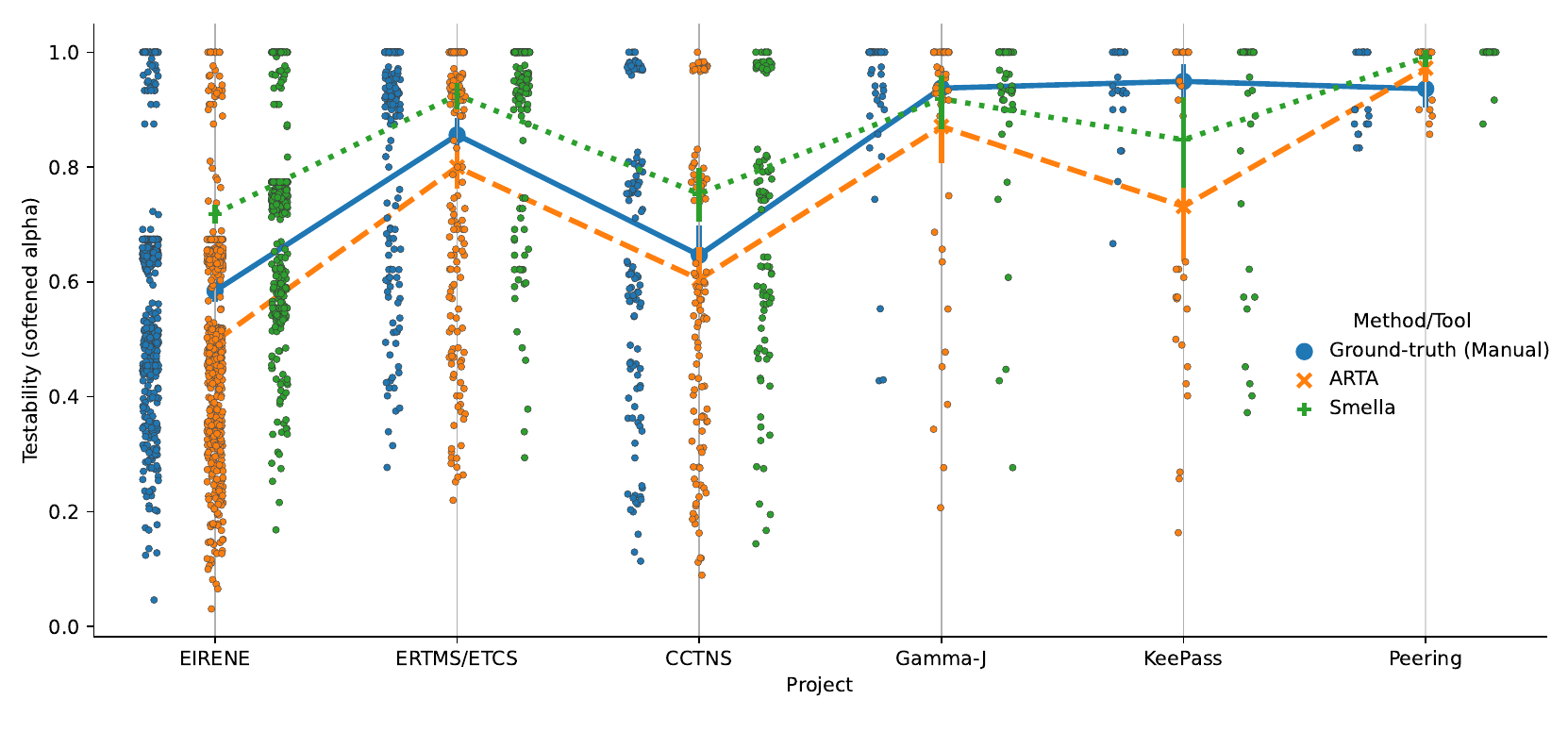}
    \caption{Comparison of requirements testability results with the ground-truth dataset (softened alpha)}
    \label{fig:requirement-testability-v4-softened}
\end{figure}

\begin{figure}[!h]
    \centering
    \includegraphics[width=0.95\linewidth]{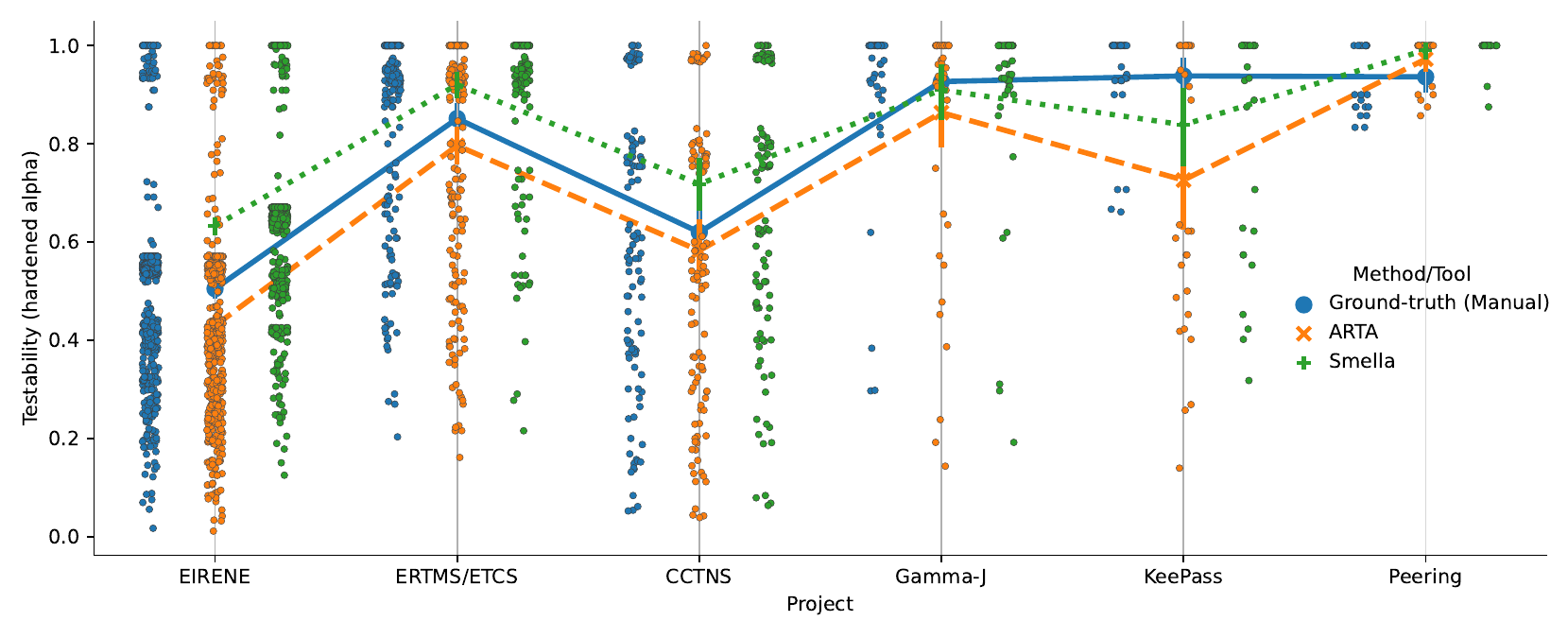}
    \caption{Comparison of requirements testability results with the ground-truth dataset (hardened alpha)}
    \label{fig:requirement-testability-v4-hardened}
\end{figure}

\noindent
 Figure \ref{fig:requirement-testability-v4-softened} depicts the mean of testability along with the standard deviation error for each project in our dataset when the softened policy is used for alpha. Figure \ref{fig:requirement-testability-v4-hardened} shows the same diagram when the hardened policy is chosen for the alpha. The testability of the individual requirements also has been shown with data points on both figures.
It can be observed that our proposed tool, ARTA, underestimates requirements testability in most (five of six) projects in our dataset, while the hand-made dictionary used in Smella often overestimates the testability.
ARTA produces testability values that are closer to the ground-truth than the ones produced by Smella \cite{Femmer20172} in four of six projects, including EIRENE, ERTMS/ETCS, CCTNS, and Peering. About 92\% of the requirements in our datasets belong to these projects. 
Figures \ref{fig:requirement-testability-v4-softened} and \ref{fig:requirement-testability-v4-hardened} also reveal that most requirements of the evaluated systems are testable since the data points are denser in the top area of the charts.
 
\begin{figure}[!h]
    \centering
    \begin{subfigure}
        {0.45\linewidth}
        \centering
        \includegraphics[width=0.95\linewidth]{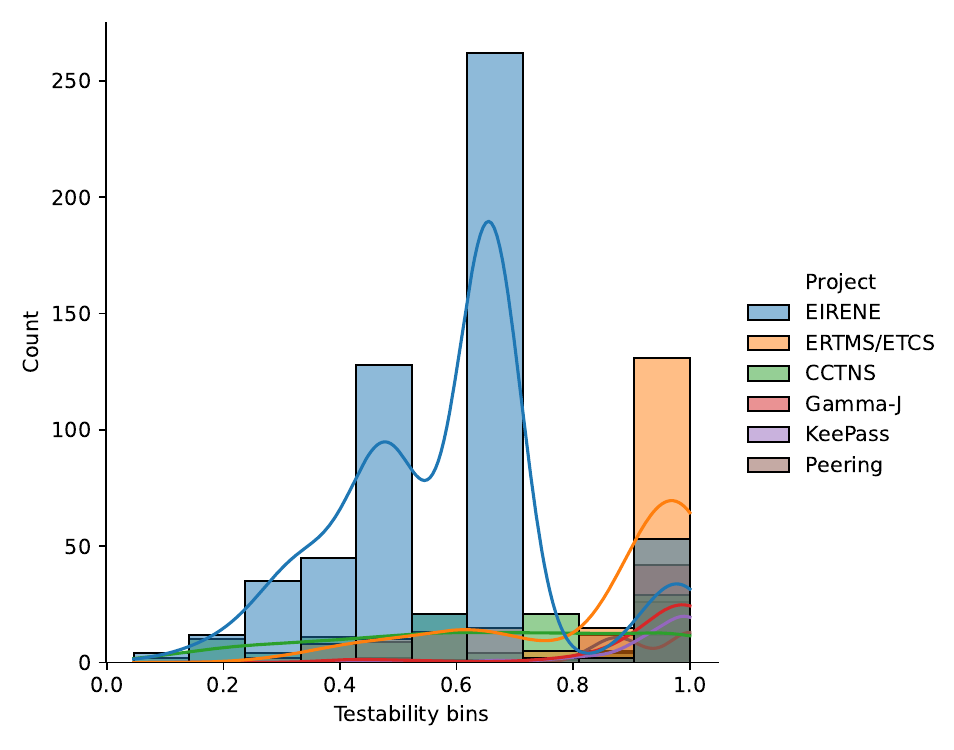}
        \caption{Ground truth testability (softened alpha)}
        \label{fig:15a}
    \end{subfigure}
\begin{subfigure}
        {0.45\linewidth}
        \centering
        \includegraphics[width=0.95\linewidth]{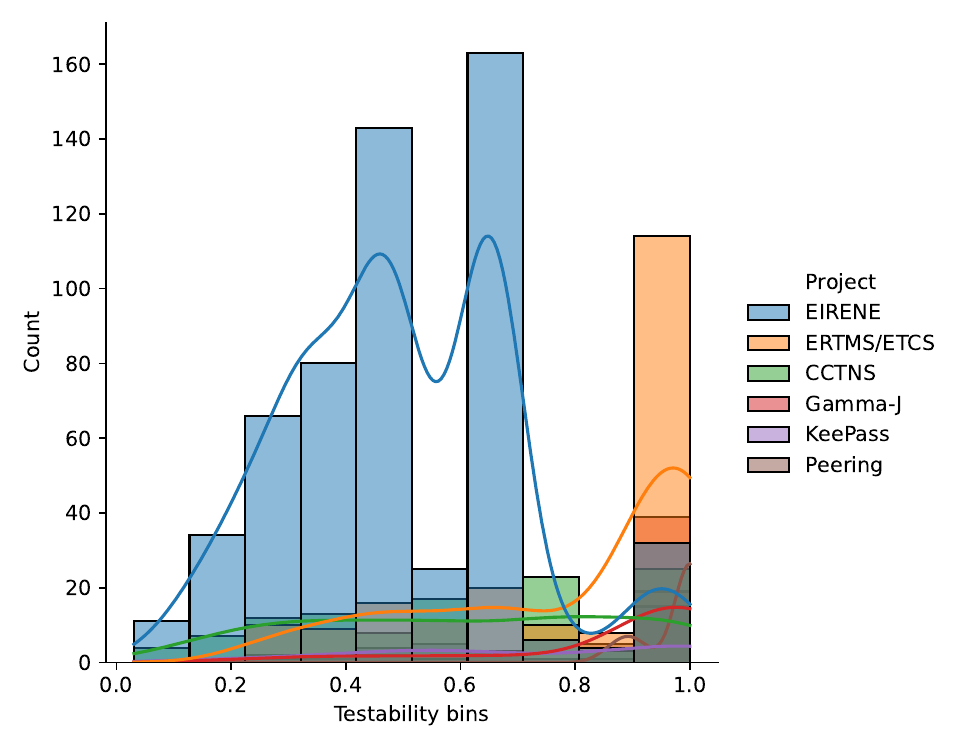}
        \caption{ARTA testability (softened alpha)}
        \label{fig:15b}
    \end{subfigure}\\
 \begin{subfigure}
    {0.45\linewidth}
    \centering
    \includegraphics[width=0.95\linewidth]{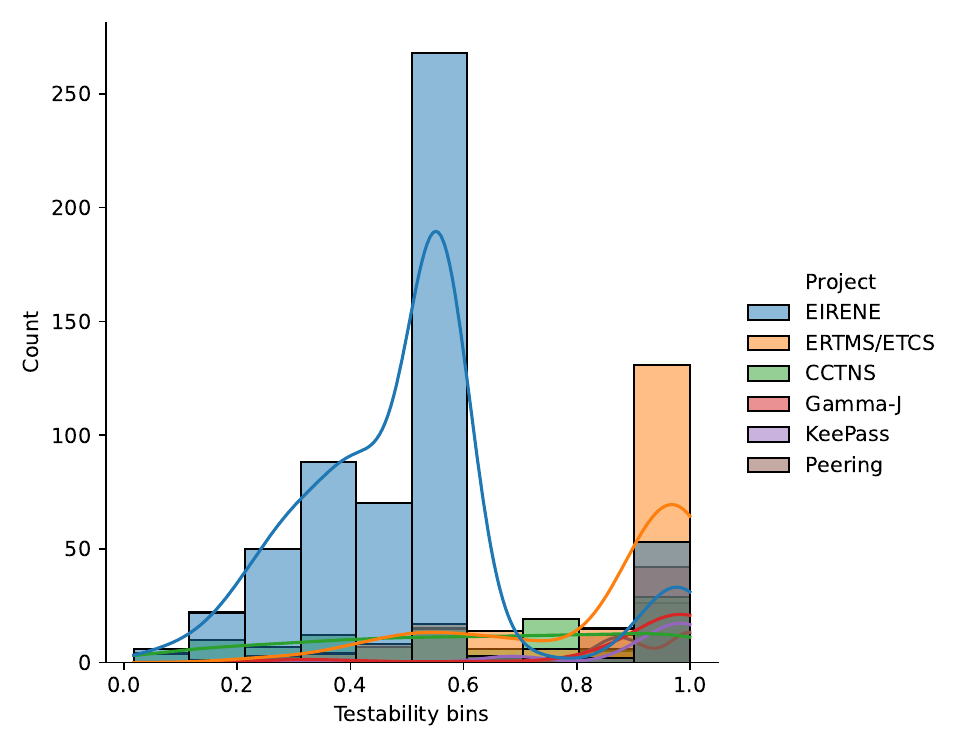}
    \caption{Ground truth testability (hardened alpha)}
    \label{fig:15c}
\end{subfigure}
\begin{subfigure}
    {0.45\linewidth}
    \centering
    \includegraphics[width=0.95\linewidth]{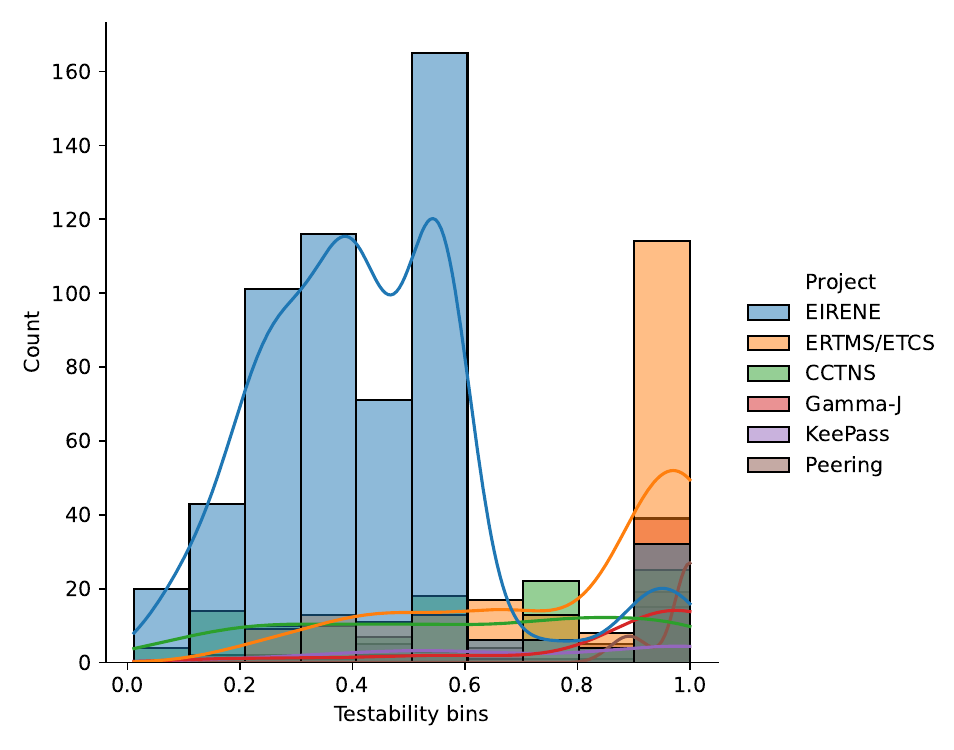}
    \caption{ARTA testability (hardened alpha)}
    \label{fig:15d}
\end{subfigure}

    \caption{Distribution of the requirements testability values}
    \label{fig:15}
\end{figure}

Figure \ref{fig:15} gives detailed information about the requirements testability status in the prepared dataset and proposed approach. It illustrates the distribution of testability values for the requirements on each project with histogram plots.
Figures \ref{fig:15a} and \ref{fig:15c} show the distribution of ground-truth values with softened alpha.
Figures \ref{fig:15b} and \ref{fig:15d} show the distribution of estimated values via our proposed method for measuring requirements testability based on smells.
Histogram plots indicate that our tool, ARTA, preserved the distribution of ground-truth values.
Indeed, the distribution of requirements testability values computed by our proposed model is highly similar to the ground-truth testability values in the prepared dataset, indicating that the proposed testability model is accurate at all testability levels (histogram bins). Moreover, requirements testability values have peaks at $[0.55,0.75]$ and $[0.90,1.0]$, revealing moderate to high testability for most requirements.

Table \ref{table:testability-measurement-error} shows the value of error metrics for estimating testability by our method with two softened and hardened policies for alpha. The total mean squared error is between 0.0293 and 0.0339 (respectively for hardened and softened alpha), which means that our approach can measure the testability of the requirements based on requirement smells with a relatively low error ($ < 5\%$). 
At the same time, error values are relatively different for each project due to the various types and frequencies of their smells. For instance, in the KeePass project, we got the worst results. Figure \ref{fig:requirement-testability-comparing-alphas} shows that 32 requirements in the KeePass project are almost clean and contain only a short sentence. Hence, the KeePass requirements are highly testable. 
Underestimating the testability of the requirements for the KeePass project originates from a relatively high false-positive rate of smells, such as vague pronouns, which improvement of them can be the subject of future works. Another important observation is that the prediction results are a bit more accurate with a hardened policy for the alpha parameter. The main reason is that the impact of test effort is increased while the impact of test quality remains fixed in terms of requirement clarity. Nevertheless, the difference between the upper and lower bound of testability measurement error is negligible, which means that our approach is stable regarding the alpha parameter.

\begin{table}
    \centering
    \caption{Testability measurement errors}
    \label{table:testability-measurement-error}
    \arrayrulecolor{black}
    \resizebox{0.70\linewidth}{!}{%
        \begin{tabular}{lllllll}
            \arrayrulecolor{black}\hline
            Project        & Alpha                                                                           & MAE                                                                         & MSE                                                                         & RMSE                                                                        & MSLnE                                                                        & MdAE                                                                         \\ 
            \arrayrulecolor{black}\hline
            EIRENE         & \begin{tabular}[c]{@{}l@{}}Softened\\Hardened\end{tabular}                      & \begin{tabular}[c]{@{}l@{}}0.1230 \\ 0.1050\end{tabular}                    & \begin{tabular}[c]{@{}l@{}}0.3001 \\0.0229\end{tabular}                     & \begin{tabular}[c]{@{}l@{}}0.1735 \\ 0.1514\end{tabular}                    & \begin{tabular}[c]{@{}l@{}}0.0135 \\0.0109\end{tabular}                      & \begin{tabular}[c]{@{}l@{}}0.1076 \\0.0899\end{tabular}                      \\
            ERTMS/ETCS     & \begin{tabular}[c]{@{}l@{}}Softened\\Hardened\end{tabular}                     & \begin{tabular}[c]{@{}l@{}}0.1246 \\ 0.1236\end{tabular}                    & \begin{tabular}[c]{@{}l@{}}0.0431 \\0.0428\end{tabular}                     & \begin{tabular}[c]{@{}l@{}}0.2076 \\0.2068\end{tabular}                     & \begin{tabular}[c]{@{}l@{}}0.0151 \\ 0.0150\end{tabular}                     & \begin{tabular}[c]{@{}l@{}}0.0588 \\0.0588\end{tabular}                      \\
            CCTNS          & \begin{tabular}[c]{@{}l@{}}Softened\\Hardened\end{tabular}                     & \begin{tabular}[c]{@{}l@{}}0.1012 \\0.0952\end{tabular}                     & \begin{tabular}[c]{@{}l@{}}0.0233 \\0.0218\end{tabular}                     & \begin{tabular}[c]{@{}l@{}}0.1525 \\0.1475\end{tabular}                     & \begin{tabular}[c]{@{}l@{}}0.0088 \\ 0.0082\end{tabular}                     & \begin{tabular}[c]{@{}l@{}}0.0449 \\0.0421\end{tabular}                      \\
            Gamma-J        & \begin{tabular}[c]{@{}l@{}}Softened\\Hardened\end{tabular}                     & \begin{tabular}[c]{@{}l@{}}0.1081 \\ 0.1044\end{tabular}                    & \begin{tabular}[c]{@{}l@{}}0.0307 \\ 0.0295\end{tabular}                    & \begin{tabular}[c]{@{}l@{}}0.1753 \\0.1718\end{tabular}                     & \begin{tabular}[c]{@{}l@{}}0.0106 \\0.0100\end{tabular}                      & \begin{tabular}[c]{@{}l@{}}0.0667 \\0.0667\end{tabular}                      \\
            KeePass        & \begin{tabular}[c]{@{}l@{}}Softened\\Hardened\end{tabular}                     & \begin{tabular}[c]{@{}l@{}}0.2326 \\0.2270\end{tabular}                     & \begin{tabular}[c]{@{}l@{}}0.1061 \\ 0.1010\end{tabular}                    & \begin{tabular}[c]{@{}l@{}}0.3257 \\0.3178\end{tabular}                     & \begin{tabular}[c]{@{}l@{}}0.0392 \\0.0376\end{tabular}                      & \begin{tabular}[c]{@{}l@{}}0.1577 \\0.1427\end{tabular}                      \\
            Peering        & \begin{tabular}[c]{@{}l@{}}Softened\\Hardened\end{tabular}                     & \begin{tabular}[c]{@{}l@{}}0.0524 \\0.0524\end{tabular}                     & \begin{tabular}[c]{@{}l@{}}0.0069 \\0.0069\end{tabular}                     & \begin{tabular}[c]{@{}l@{}}0.0828 \\0.0828\end{tabular}                     & \begin{tabular}[c]{@{}l@{}}0.0018 \\0.0018\end{tabular}                      & \begin{tabular}[c]{@{}l@{}}0.0 \\ 0.0\end{tabular}                               \\
            \textbf{All requirements} & \begin{tabular}[c]{@{}l@{}}\textbf{Softened}\\\textbf{Hardened}\end{tabular} & \begin{tabular}[c]{@{}l@{}}\textbf{0.1218}\\\textbf{0.1102}\end{tabular} & \begin{tabular}[c]{@{}l@{}}\textbf{0.0339}\\\textbf{0.0293}\end{tabular} & \begin{tabular}[c]{@{}l@{}}\textbf{0.1840}\\\textbf{0.1711}\end{tabular} & \begin{tabular}[c]{@{}l@{}}\textbf{0.0137}\\\textbf{0.0120}\end{tabular} & \begin{tabular}[c]{@{}l@{}}\textbf{0.0795}\\\textbf{0.0714}\end{tabular}  \\
            \hline
        \end{tabular}
    }
    \arrayrulecolor{black}
\end{table}

\subsubsection{Evaluation by experts}
\noindent
To evaluate the accuracy and usefulness of the proposed testability measurement formula, we randomly selected 21 requirements from each project and ranked them in the order of their testability magnitude, computed by Equation \ref{eq:4}. Thereafter, we asked three independent experts, highly experienced in requirements engineering, to manually score the selected requirements based on the extent to which objectives and feasible tests can be designed to determine whether each requirements is met or not \cite{ISO/IEC/IEEE20171}. 
Our experts were asked to score the requirements testability between 1 and 10 for each selected requirement. Then, we averaged the testability scores assigned by the experts to each requirement and computed the rank of manual evaluation for each requirement. Finally, we computed the Spearman rank-order correlation between the requirements’ ranks provided by our approach and the ranks by the manual evaluation scores. 
    
Table \ref{table:ranked-correlation} shows the correlation analysis between our ranking results and the experts’ opinions about the requirements. The last row of the table shows the results for all requirements. A correlation coefficient of 0.93 is observed, for all evaluated requirements in this experiment in the last row. The null hypothesis that the result occurred by chance is rejected with a p-value less than 0.01.

The correlation coefficient for all evaluated requirements is close to one ($\approx$0.93), indicating that our tool can rank the requirement statements based on their testability values such that the ranks are in high agreement with the ones obtained by experienced requirement engineers. This way, our tool can measure the requirement testability automatically. The ideal value for the correlation coefficient is one, which means there is no difference between manual and automated ranking. It is worth noting that the p-value for the 0.93 correlation coefficient is less than 0.01, indicating a confidence level of 99\% for the obtained correlation coefficient. Therefore, the correlation coefficient of 0.93 confirms the accuracy and usefulness of our requirements testability measurement approach.

\begin{table}
    \centering
    \caption{Spearman rank-order correlation coefficient with the associated p-value between expert testability scores and ARTA}
    \label{table:ranked-correlation}
    \arrayrulecolor{black}
    \resizebox{0.75\linewidth}{!}{%
        \begin{tabular}{lllll} 
            \hline
            \rowcolor[rgb]{0.949,0.949,0.949} {\cellcolor[rgb]{0.949,0.949,0.949}}                          & \multicolumn{2}{l}{Softened
                value for testability} & \multicolumn{2}{l}{Hardened
                value for testability}  \\ 
            \hhline{>{\arrayrulecolor[rgb]{0.949,0.949,0.949}}->{\arrayrulecolor{black}}----}
            \rowcolor[rgb]{0.949,0.949,0.949} \multirow{-2}{*}{{\cellcolor[rgb]{0.949,0.949,0.949}}Project} & Spearman correlation & P-value                       & Spearman correlation & P-value                        \\ 
            \hline
            EIRENE                                                                                          & 0.9547               & 6.3744$\times 10^{-11}$                  & 0.9349               & 1.5555$\times 10^{-9}$                    \\
            ERTMS/ETCS                                                                                      & 0.8590               & 1.2358$\times 10^{-6}$                   & 0.8590               & 1.2358$\times 10^{-6}$                    \\
            CCTNS                                                                                           & 0.9324               & 2.1803$\times 10^{-9}$                   & 0.9423               & 5.4530$\times 10^{-10}$                   \\
            Gamma-J                                                                                         & 0.8749               & 4.4837$\times 10^{-7}$                   & 0.8749               & 4.4837$\times 10^{-7}$                    \\
            KeePass                                                                                         & 0.8900               & 1.4816$\times 10^{-7}$                   & 0.8973               & 8.2228$\times 10^{-8}$                    \\
            Peering                                                                                         & 0.5399               & $1.3910 \times 10^{-2}$                        & 0.5399               & $1.3990 \times 10^{-2}$                         \\
            \rowcolor[rgb]{0.949,0.949,0.949} Total                                                         & 0.9310               & 1.7530$\times 10^{-53}$                  & 0.9321               & 6.5681$\times 10^{-8}$                    \\
            \hline
        \end{tabular}
    }
\end{table}

\subsubsection{Qualitative analysis}\label{sec:qualitative-analysis}
\noindent
To make sense of the concept of requirements testability based on our proposed method, several concrete examples selected from our requirement dataset, along with their smells, clarity, and testability, are shown in Table \ref{table:qualitative-analysis}. 
Actual smells have been \textbf{bolded} throughout the text, and words identified by ARTA as smells have been marked with an \underline{underline}. Sentences within a requirement are separated by $\bullet$ symbol.  
The upper bound and lower bound of testability have been reported considering softened and hardened policies for alpha. Testability has also been computed based on our automatic method and shown inside parentheses underneath the ground-truth values.

 Focusing on Table \ref{table:qualitative-analysis}, we observe some exciting results supporting the applicability of our proposed testability measurement mathematical model. For example, the clarity of both the requirements R1 and R2 is almost equal, while it is evident that testing and validating R1, presumably, requires more effort than R2. The effort is taken into account by both the alpha parameter and the number of sentences in the requirement. Our testability quantification model can correctly rank these two requirements. On the other hand, the number of sentences of both the requirements R2 and R3 are equal. However, the clarity of R2 is more than R3. Therefore, the testability of R3 is less than R2. 
 
 The model has correctly detected all the smells in R1 and R2 definitions that imply no testability measurement error. However, in requirement R3, our model has detected the word "can" as a requirement smell incorrectly. The reason is that sometimes, and not all the time, "can" is used in the sense of "may" to give permission. It needs a large corpus of texts labeled with the definition of "can" as "can" or "may" to use Word2Vec analysis for learning the contexts in which "can" can be replaced with "may."
 
 Requirement R4 shows another desired aspect of our testability model. This requirement contains eight smells of six different types, making it too smelly. However, all smells have occurred in one sentence, \emph{i.e.}, the minimum possible unit for declaring a requirement in natural language. Therefore, there is no cost for extra sentences when testing the requirement. Indeed, it only expresses one requirement.
It is observed that both the requirement smells and the requirement's length are indicators of the requirements testability.
Table \ref{table:qualitative-analysis} also denotes that there are various requirements with different levels of testability in our labeled dataset. Therefore, it can be used in future works on requirement quality analysis.

 \begin{table}
     \centering
     \caption{Examples of requirements with smells and their clarity and testability values}
     \label{table:qualitative-analysis}
     \resizebox{1.0\linewidth}{!}{%
         \begin{tabular}{lllll} 
             \hline
             \multirow{2}{*}{Requirement}        & \multirow{2}{*}{Project} & \multirow{2}{*}{Requirement 
                 clarity} & \multicolumn{2}{l}{
                 Requirement testability
             }   \\  \cline{4-5}
             &                          &                                          & Softened alpha                                         & Hardened alpha                                          \\ 
             \hline
             \begin{tabular}[c]{@{}l@{}}R1: "For \uline{\textbf{calls}} between a controller and the lead cab, it shall be possible \\to add the controller to the multi-driver \uline{\textbf{call}} $\bullet$ Either the lead driver \uline{\textbf{calls}} \\the controller or the controller \uline{\textbf{calls}} the lead driver $\bullet$ In the latter \uline{\textbf{case}}, \\the controller is automatically added into the multi-driver \textbf{\uline{call}}  $\bullet$ Functional \\identity of the controller shall be displayed in the leading cab $\bullet$"\\  \end{tabular} & EIRENE                   & 0.69                                     & \begin{tabular}[c]{@{}l@{}}0.21 \\(0.21)\end{tabular}  & \begin{tabular}[c]{@{}l@{}}0.13 \\ (0.13) \end{tabular}  \\ 
             \hline
             \begin{tabular}[c]{@{}l@{}}R2: "The composition of \textbf{\uline{call}} groups shall be able to be modified within \\the network $\bullet$ A single-user shall be able to be a member of one or \uline{\textbf{more}} \uline{\textbf{}}\\\uline{\textbf{call}} groups$\bullet$"\\  \end{tabular}                                                                                                                                                                                                                                                                                & EIRENE                   & 0.68                                     & \begin{tabular}[c]{@{}l@{}}0.46 \\(0.46)\end{tabular}  & \begin{tabular}[c]{@{}l@{}}0.39 \\(0.39)\end{tabular}   \\ 
             \hline
             \begin{tabular}[c]{@{}l@{}}R3: "The maximum length of a message \textbf{segment} shall be 96 characters 
                 $\bullet$ \\A message \uline{can} include \uline{\textbf{several }}\textbf{segments} $\bullet$"\\  \end{tabular}                                                                                                                                                                                                                                                                                                                                                          & EIRENE                   & 0.58                                     & \begin{tabular}[c]{@{}l@{}}0.39 \\(0.44)\end{tabular}  & \begin{tabular}[c]{@{}l@{}}0.33 \\(0.37)\end{tabular}   \\ 
             \hline
             \begin{tabular}[c]{@{}l@{}}R4: "If a document is either \textbf{too} \uline{\textbf{long}}, dispersed over \uline{\textbf{several}} \uline{\textbf{pages}} or in \\a \textbf{specific} layout \uline{\textbf{that}} is \uline{not} \textbf{suitable} for online reading, a printer-friendly \\version of the document \uline{\textbf{should}} be provided \uline{that} prints the content in a \uline{}\\\uline{form} acceptable to the user (\emph{e.g.} in the expected layout, paper format, or \\orientation) $\bullet$"\\  \end{tabular}                                            & CCTNS                    & 0.27                                     & \begin{tabular}[c]{@{}l@{}}0.27 \\(0.27)\end{tabular}  & \begin{tabular}[c]{@{}l@{}}0.27 \\ (0.27)\end{tabular}  \\ 
             \hline
             \begin{tabular}[c]{@{}l@{}}R5: "When the train is stationary or after a \uline{\textbf{certain}} time (\emph{e.g.} the time for \\"route releasing" of the overlap, the \textbf{release} speed calculation shall be \\based on the distance to the \uline{\textbf{danger}} \uline{\textbf{point}} (if calculated on-board) $\bullet$ The \\condition for \uline{this} change shall be defined for each target as infrastructure\\ data $\bullet$"\\  \end{tabular}                                                                                                                       & ERTMS/
             ETCS            & 0.72                                     & \begin{tabular}[c]{@{}l@{}}0.45 \\(0.35)\end{tabular}  & \begin{tabular}[c]{@{}l@{}}0.38 \\ (0.30)\end{tabular}  \\ 
             \hline
             \begin{tabular}[c]{@{}l@{}}R6: "All data related to the \textbf{word} must be shown $\bullet$ For example, if user\\ \uline{\textbf{types}} "abc" and abc is \uline{part} of a password and a username, both entries \\must be shown $\bullet$"\\  \end{tabular}                                                                                                                                                                                                                                                                                                                  & KeePass                  & 0.93                                     & \begin{tabular}[c]{@{}l@{}}0.77 \\ (0.57)\end{tabular} & \begin{tabular}[c]{@{}l@{}}0.66 \\ (0.48)\end{tabular}  \\ 
             \hline
             \begin{tabular}[c]{@{}l@{}}R7: "The system will employ on demand asynchronous loading for \textbf{\uline{faster}} \\execution of \textbf{\uline{pages}}$\bullet$"\\   \end{tabular}                                                                                                                                                                                                                                                                                                                                                                                               & Gamma-J                  & 0.61                                     & \begin{tabular}[c]{@{}l@{}}0.61 \\(0.61)\end{tabular}  & \begin{tabular}[c]{@{}l@{}}0.61 \\ (0.61)\end{tabular}  \\ 
             \hline
             \begin{tabular}[c]{@{}l@{}}R8: "Primary CDN has already acquired \textbf{sufficient} \uline{\textbf{external}} resources $\bullet$"
\end{tabular}  & Peering                  & 0.50                                     & \begin{tabular}[c]{@{}l@{}}0.50 \\(0.88)\end{tabular}  & \begin{tabular}[c]{@{}l@{}}0.50 \\ (0.88)\end{tabular}  \\
             \hline
         \end{tabular}
     }
 \end{table}

\subsubsection{Important smells affecting requirements testability}
\noindent
To understand which smells are more important in determining requirements testability, we fit a decision tree regressor \cite{scikit-learn2011} on our dataset using the number of each smell as independent variables or features and the value of testability as dependent variables. The decision tree regressor, at each step, finds the best feature and divides samples based on the most appropriate value of this feature, such that the regression error, \emph{e.g.}, mean squared error (MSE), is minimized. 

\begin{figure}
    \centering
    \includegraphics[width=0.95\linewidth]{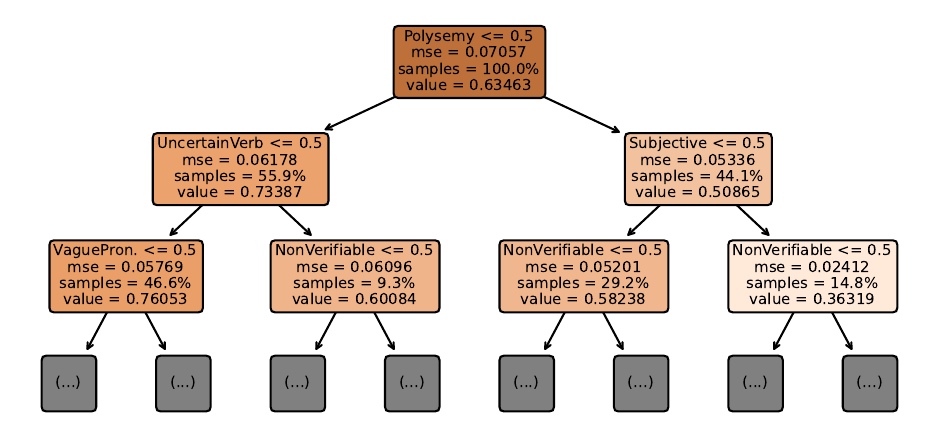}
    \caption{Most important smells affecting requirements testability}
    \label{fig:16}
\end{figure}

Figure \ref{fig:16} shows three top levels of the decision tree regressor after training on the dataset containing the lower bounds of testability values obtained by our proposed method.
 The selected feature, its cut-off value, \emph{i.e.}, the value of the feature in which data is divided into the separated sub-trees, the mean squared error (MSE) of division, the portion of samples, and the Gini index as impurity value is shown on each node. 
The root of the decision tree regressor indicates that the most important feature to estimate testability is the polysemy smell with a cut-off point of $0.5$, MSE of $0.0706$, and impurity of $0.6346$. Uncertain verbs and subjective language constitute the second level of decision tree regressor. 
It is observed that frequent requirement smells, \emph{e.g.}, polysemy, have more negative impacts on testability than other smells. However, a dataset with many labeled samples is still required to build a more reliable machine-learning model to predict testability.
It is worth noting that distinguishing testable requirements from non-testable ones is not a trivial machine learning task since most requirements are testable, and therefore, the dataset is imbalanced.

\begin{tcolorbox}
    \textbf{Answer to RQ\textsubscript{4}}:
    \emph{
        Requirement testability could be determined based on requirement smell and length of the requirement with a mean absolute error of $0.12$ and mean squared error of $0.03$. The most important smells affecting requirements testability are polysemy, uncertain verbs, and subjective language. Prevalent requirement smells affect requirements testability more than other smells.
    }
\end{tcolorbox}

\section{Threats to validity}\label{sec:threats-to-validity}
\noindent
We argue the most important threats that may affect the validation of our proposed method and its evaluation and discuss how we mitigate them. 

\subsection{Threats to construct validity}\label{sec:threats-to-contruct-validity}
\noindent
The main threat to the construct validity of our approach is the proposed testability formula. Our requirements testability formula aims to provide a means to \emph{compare and rank} requirements based on their testability measures. It is based on simple heuristics to measure requirements testability from the viewpoint of the requirement engineer obtained by applying the goal-question-metric (GQM) approach \cite{Basili1994}. The dissimilarities of various domains with computer science have been measured empirically by applying  a word embedding technique on a considerable corpus of Wikipedia documents. Other aspects affecting the parameter alpha are ordinal according to the importance of different levels in each aspect. For instance, the cost of acceptance testing of a requirement in safety-critical systems is more than a non-critical system \cite{Dubrova2013}. Further empirical validations must be performed to tune the numerical parameters, requiring large annotated requirement smell datasets.

\subsection{Threats to internal validity}\label{sec:threats-to-internal-validity}
\noindent
Threats to internal validity are related to the correctness of the experiment outcomes. A major threat to the internal validity of our approach results is that the experience of our postgraduate students, as well as the experts, could play a role in their judgments on the detection of smells in our requirements dataset. As described in Section \ref{sec:case-studies}, initially, 5,000 requirements were labeled by the students. 
The requirements were discussed and labeled by four groups of three postgraduate students. Initially, we used these 5,000 labeled requirements to evaluate our smell detection approach. When comparing the manually detected smells with the ones detected by our approach, we came across relatively high false negative and false positive rates for some types of smells. 

In some cases, the difficulty was with our approach, and we corrected the approach as much as we could. However, we noticed that in most cases, the requirements were not labeled appropriately. 
Thereafter, the manually labeled dataset provided by our students was revised several times by experts in the software industry. We mitigated the threat by discarding those requirements for which there was not a total agreement on the type of smells. Finally, we ended up with 985 requirements that both the experts in software industries and our teams agreed on their types of smells. 
There are still false negatives and false positives when comparing the smells detected by our approach with the ones detected manually and included in our data set.

Another internal threat is the choice of word-embedding algorithm and hyperparameters used to configure the algorithm. Different word-embedding algorithms can be used to train Word2Vec models \cite{Mikolov2013efficient, Pennington2014}. 
We used the Continuous Bag of Words (CBOW) algorithm \cite{Mikolov2013efficient}, one of the state-of-the-art neural word-embedding algorithms, with default hyperparameters available in the Gensim library \cite{Gensim2020}. 
According to Mikolov et al. \cite{Mikolov2013efficient} the CBOW Word2Vec algorithm is several times faster to train than the well-known Skip-gram Word2Vec algorithm. 
The computational complexity of CBOW depends on the vocabulary size ($V$), the context window size ($C$), the embedding dimension ($D$), and the hidden layer size ($H$). The number of operations per training iteration is $O(N\times(D\times C+D\times H+H\times V))$, where $N$ is the number of training examples. The complexity can be reduced by using techniques that decrease $V$ or $N$, such as hierarchical softmax, negative sampling, or subsampling. These techniques make CBOW  \cite{Mikolov2013efficient} more efficient and scalable without affecting the word embedding quality.
In addition, CBOW is slightly more accurate than other Word2Vec techniques, such as Skip-gram for the frequent words \cite{Mikolov2013efficient, Giatsoglou2017}, making it more suitable for building our smelly words dictionary. Finally, the CBOW architecture works better than the neural net language models (NLMs) on the semantic-syntactic relationship task \cite{Mikolov2013efficient}.
For these reasons, we chose CBOW to find the most frequent words in computer science, which may contain different meanings in other domains. Nevertheless, exploring other word-embedding models and their suit hyperparameters for building the smelly word dictionary is part of our ongoing work.

\subsection{Threats to external validity}\label{sec:threats-to-external-validity}
\noindent
Threats to external validity are related to the generalization of our outcomes. The outcomes might be affected by three factors. Firstly, our dictionary of smelly words is limited to only ten different application domains. Therefore, it could be of no use for the other domains. However, the dictionary can be extended automatically with the smelly words in any domain. Secondly, we have considered only nine requirement smells to compute our proposed requirement evaluation metric which is called requirement clarity. Newly introduced smells should be integrated with our tools.

The third external threat is that our experiments have been performed on natural language requirements in English. Therefore, the precision and recall value may differ for requirements written in other languages. However, our dictionary-building approach can be applied to application domains in any language since Wikipedia articles are available for a wide range of human languages. 

\section{Conclusion}\label{sec:conclusion}
\noindent
Software requirements testability can be expressed and measured as a function of requirement smells and the requirement's length. In this paper, we show that a newly defined requirement smell, Polysemy, along with Subjective language and Non-verifiable terms, are the most prevalent smells in the software requirements, resulting in non-testable requirements. These smells could be detected using a comprehensive dictionary provided by training a Word2Vec model on a massive textual corpus in different domains and ranking ambiguous words and terms with many possible meanings. 
The proposed dictionary is extended to include smelly words rather than solely nouns in ten different application domains. The experiments with nearly 1,000 software requirements from six well-known industrial and academic projects demonstrate that the proposed method supported by our web-based tool, ARTA, outperforms Smella, a state-of-the-art tool, to detect requirement smells. ARTA improves the Precision, Recall, and F1 score of dictionary-based smell detection, respectively, with an average of 0.0311, 0.3299, and 0.2824. As a result, ARTA measures the requirement testability with a mean absolute error of 0.12 and a mean squared error of 0.03. The correlation coefficient between the requirements testability ranks computed by ARTA and the ranks obtained by experts’ opinions is close to one indicating the applicability of the proposed approach for use in practice.

Our results show that while only 4.93\% of words are recognized as smelly words in the software requirements, 72.39\% of requirements contain at least one kind of smell. The frequency of requirements smells in each requirement is similar to the code smells. Subjective languages and Polysemy are among the most crucial smells affecting requirements testability, and the number of smells in the requirements increases as the requirement length grows. Therefore, both the requirements smells and the length of the requirements must be considered when measuring the testability of requirements definitions. Quantifying requirements testability in terms of requirement smells allows requirement engineers and domain experts to know about the issues in requirements with the exact location of the issue and refactor the smelly parts as soon as possible. It directly affects the performance of functional and acceptance testing in the software development lifecycle.

Future works can focus on utilizing the ARTA platform to enlarge the dataset for more requirements and then use machine learning predictive models to predict requirements smells as well as requirements testability based on the requirement statement. Most smells are identified by analyzing the whole sentence rather than a single word. We plan to improve smell detection performance, specifically reducing the false-positive rate by applying Recurrent Neural Networks (RNNs) \cite{Goodfellow2016} and Transformer models \cite{Vaswani2017}. RNNs and Transformers are new deep learning models that can be used to consider the entire requirement text for identifying smells without any dictionary or word-level analysis.
Another opportunity for the study is to propose a method for automatically refactoring the requirement smells after detection. Similar methods in NLP tasks, such as anaphora resolution, which most commonly appears as pronoun resolution, can be exploited as primary solutions for recommending a proper word to use instead of smelly words.

\section*{Declarations}
\subsection*{Data Availability Statement}
\noindent The datasets generated and analyzed during the current study are available in Zenodo, \href{https://doi.org/10.5281/zenodo.4266727}{\emph{https://doi.org/10.5281/zenodo.4266727}}.

\subsection*{Funding}
\noindent This study has received no funding from any organization. 

\subsection*{Conflict of Interest}
\noindent All of the authors declare that they have no conflict of interest. 

\subsection*{Ethical Approval}
\noindent This article does not contain any studies with human participants or animals performed by any of the authors.


\bibliography{bibliographydb.bib}

\end{document}